# Interval-valued q-rung orthopair fuzzy Weber operator and its group decision-making application

**Benting Wan[a], Zhuocheng Wu[a,*], Mengjie Han[b] , Minjun Wan[a]**

[a]*Jiangxi University of Finance and Economics, Nanchang, 330013, China*

[b]*School of Information and Engineering, Dalarna University, Falun, 79188, Sweden*

ORCID 0000-0001-6643-2971, ORCID 0009-0004-6881-9247, ORCID 0000-0003-4212-8582,
ORCID 0009-0000-7793-4723

*e-mail*: *wanbenting@jxufe.edu.cn*,**e-mail: 2202004222@stu.jxufe.edu.cn* ,

*e-mail*：*mea@du.se, e-mail: 2202220770@stu.jxufe.edu.cn*

**Abstract:** The evaluation of learning effectiveness requires the integration of objective test results and analysis of uncertain subjective evaluations. Fuzzy theory methods are suitable for handling fuzzy information and uncertainty to obtain comprehensive and accurate evaluation results. In this paper, we develop a Swing-based multi-attribute group decision-making (MAGDM) method under interval-valued q-rung orthopair fuzzy sets (IVq-ROFSs). Firstly, an extended interval-valued q-rung orthopair Weber ordered weighted average (IVq-ROFWOWA) operator is introduced. Then the attribute weights deriving method is designed by using the optimized Swing algorithm. Furthermore, we develop a MAGDM method for evaluating students' learning effectiveness using the IVq-ROFWOWA operator and the Swing algorithm. Finally, a case of evaluating students' learning effectiveness is illustrated by using the proposed MAGDM method. The implementing results demonstrate that the proposed MAGDM method is feasible and effective, and the Swing algorithm enhances better differentiation in ranking alternatives compared to other methods.

## 1 Introduction

Learning effectiveness evaluation is a dynamic, uncertain, and complex task with subjective and objective elements. Learning effectiveness evaluation can effectively express learners' status and efficiency, thus providing facilitation services for learners (Araghi et al., 2023；Paśko et al. ,2022; Culver, 2022). As a result, learning effectiveness evaluation has received attention from researchers. For example, Benő et al. proposed an online diagnostic learning evaluation mode relying on artificial intelligence (Csapó and Molnár, 2019). Alamri (2021) offered a learning evaluation technology model to guide students' active learning. Zhou presented a deep learning-based classroom evaluation model (Zhou et al., 2022), which can reflect students' learning status promptly and effectively. Evaluation of learning effectiveness needs to consider multiple factors of the learner. In this paper, we adopt learning interest, learning style, personality attributes, cognitive ability, and external environment as factors proposed by Wang and Lv (2022) to evaluate learning effectiveness. Fuzzy mathematics can effectively handle information of uncertainty and fuzziness in many practical applications (Laktionov et al.,2023; Wu et al., 2023). Due to the existence of uncertainty and fuzziness in this process, fuzzy intelligence method is used. One of the contributions is that we construct a Swing-based multi-attribute group decision-making (MAGDM) method under IVq-ROFSs for evaluating students' learning effectiveness.

For IVq-ROFSs, the representation of the assessor's evaluation information provides greater flexibility for the assessor. IVq-ROFS, proposed by Joshi et al. (2018), is an extension of interval-valued intuitionistic fuzzy sets (Atanassov, 1999) and q-rung orthopair fuzzy sets (Yager, 2016). The use of IVq-ROFSs can reduce the time of preparing and processing data for decision-making (Wang et al., 2019; Luqman and Shahzadi ,2023 ; Wan et al., 2023; Seker et al., 2023), which is widely used in multi-attribute decision-making (MADM) and multi-attribute group decision-making (MAGDM) problems. However, fuzzy operators are required to fuse decision makers (DMs) and attribute information (Zhong et al., 2023; Ye et al., 2023; Javed et al., 2022) for solving MAGDM problems. Many operators have been studied, such as the Frank operator (Frank, 1979; Riaz et al., 2022), the Hamacher operator (Hamacher, 1975; Darko and Liang, 2020; Asif et al., 2023), the ordered weighted average operator (OWA) proposed by Yager (1988), and the Weber operator submitted by Weber (1983). Among the Weber operators are the t-conorm sum operation with parameters and t-norm product operation. The decision maker can adjust the Weber parameters according to different data processing situations. Thus, the Weber operator has an excellent deflating ability, but it has yet to be extended to IVq-ROFSs, which will be addressed in this paper. The operator of the interval-valued q-rung orthopair Weber (IVq-ROFW) and the interval-valued q-rung orthopair Weber ordered weighted average (IVq-ROFWOWA) operator is proposed to better harnessing its strengths. And DMs could have a broader range of options under IVq-ROFSs for aggregating information.

Meanwhile, the weights of other attributes have variability because of dynamic environmental. For decision-making problems with unknown attribute weights, many methods have been developed to derive attribute weights. For example, the MABAC (Verma, 2021) is based on a step-by-step comparison of multiple attributes' importance. It calculates the boundary approximation area to quantify the comparison results and provides an effective solution for attribute weights. The Projection (Verma, 2021) can also be used to derive attribute weights based on the maximization or minimization projection of attribute values, thereby considering the importance and influence of different attributes. However, when there are associated relationships for attributes, the MABAC and projection methods are not able to deal with this situation. Inspired by the idea of Yang (Yang et al., 2020), who proposes Swing algorithm, we improve and optimize the Swing algorithm in this paper to transact the correlation attribute weights. It is based on the internal structure of bipartite graphs to construct association graphs for building substitution relationships. It not only captures the association relationships by gradually adjusting the scores of different schemes but also creates a more stable structure for the association graph construction process and enhance its ability to resist noise. Additionally, competition theory (Kvålseth, 2022) is introduced in this paper to compare the derivation of different attribute weight methods. The competition theory takes the Herfindahl-Hirschman Index (HHI) to assist analysts in evaluating the intensity and concentration of competition among different alternatives. A higher value indicates lower competition and higher engagement.

To the best of our knowledge, the competition theory has not yet been studied by applying the Weber operator in the IVq-ROFSs. Similarly, only a few approaches that utilize collaborative filtering algorithms to calculate attribute weights are found at this stage. Additionally, there needs to be more in evaluating how to derive attribute weights. Our motivations for addressing these limitations include the following:

(1) Weber operator (Weber, 1983) has been developed under fuzzy sets. However, it has not yet been developed under the IVq-ROFSs.

(2) For attributes with associated relationships, deriving the attribute weights method needs to be explored. We will explore the potential of collaborative filtering algorithms in calculating attribute weights.

(3) There are different methods for deriving attribute weights. However, how to select the optimal attribute weights method should be further developed. We will strive to put forward the evaluating efficiency of the attribute weights method.

Combining questionnaire survey data, it is used in an implementation case of the learning evaluation model. The main contributions of the papers are as follows:

(1) The IVq-ROFW and IVq-ROFWOWA operators are extended under IVq-ROFSs. Further, we study its idempotency, commutativity, monotonicity, and boundedness.

(2) We improve and optimize the Swing algorithm, which has better expressiveness compared with the MABAC (Verma, 2021) and the Projection (Zhang et al., 2019) by competition theory (Kvålseth, 2022).

(3) A novel MAGDM is constructed using the IVq-ROFWOWA operator and the Swing method.

(4) An implementation case of the learning evaluation is used to illustrate the application of the proposed Swing-based MAGDM method.

The remainder of this paper is arranged as follows. Section 2 introduces the preliminaries. Section 3 presents the IVq-ROFWOWA operator and its nature theorem. Additionally, Section 3 develops a MAGDM method based on the IVq-ROFWOWA operator and optimizes the Swing method. Section 4 implements a case relying on our Swing-based MAGDM method. Section 5 gives the conclusion and future research.

# 2 Preliminaries

## 2.1 IVq-ROFs

**Definition 2.1**(**Joshi et al., 2018**). Given the domain of discourse $X$, an IVq-ROFSs $A$ in $X$ is defined as:

$$A = \{\langle x, \mu_a(x), \nu_a(x)\rangle | x \epsilon X\} \tag{1}$$

where, the membership function $\mu_a(x)$ and non-membership function $\nu_a(x)$ are the mappings of interval value satisfying $\mu_a(x) = [\mu_a^-(x), \mu_a^+(x)] \subseteq [0,1]$ and $\nu_a(x) = [\nu_a^-(x), \nu_a^+(x)] \subseteq [0,1]$ for $0 \leq (\mu_a^+(x))^q + (\nu_a^+(x))^q \leq 1, q \geq 1$. The hesitation of $A$ is

$$\pi_A(x) = [\pi_A^-(x), \pi_A^-(x)] = \left[\sqrt[q]{1-(\mu_a^+(x))^q-(\nu_a^+(x))^q}, \sqrt[q]{1-(\mu_a^-(x))^q-(\nu_a^-(x))^q}\right]. \quad (2)$$

**Definition 2.2 (Joshi et al., 2018).** Let $a = ([\mu^-, \mu^+,], [\nu^-, \nu^+])$, $a_1 = ([\mu_{a_1}^-, \mu_{a_1}^+], [\nu_{a_1}^-, \nu_{a_1}^+])$ and $a_2 = ([\mu_{a_2}^-, \mu_{a_2}^+], [\nu_{a_2}^-, \nu_{a_2}^+])$ be three interval-valued q-rung orthopair fuzzy numbers (IVq-ROFNs) with $q \geq 1$, the basic operations of IVq-ROFNs can be defined as:

$$a_1 \oplus a_2 = \langle \begin{matrix} \left[\sqrt[q]{(\mu_{a_1}^-)^q+(\mu_{a_2}^-)^q-(\mu_{a_1}^-)^q(\mu_{a_2}^-)^q}, \sqrt[q]{(\mu_{a_1}^+)^q+(\mu_{a_2}^+)^q-(\mu_{a_1}^+)^q(\mu_{a_2}^+)^q}\right], \\ [\nu_{a_1}^-\nu_{a_2}^-, \nu_{a_1}^+\nu_{a_2}^+] \end{matrix} \rangle, \quad (3)$$

$$a_1 \otimes a_2 = \langle \begin{matrix} [\mu_{a_1}^-\mu_{a_2}^-, \mu_{a_1}^+\mu_{a_2}^+], \\ \left[\sqrt[q]{(\nu_{a_1}^-)^q+(\nu_{a_2}^-)^q-(\nu_{a_1}^-)^q(\nu_{a_2}^-)^q}, \sqrt[q]{(\nu_{a_1}^+)^q+(\nu_{a_2}^+)^q-(\nu_{a_1}^+)^q(\nu_{a_2}^+)^q}\right] \end{matrix} \rangle, \quad (4)$$

$$\lambda a = \langle \left[\sqrt[q]{1-(1-(\mu^-)^q)^\lambda}, \sqrt[q]{1-(1-(\mu^+)^q)^\lambda}\right], [(\nu^-)^\lambda, (\nu^+)^\lambda] \rangle, \quad (5)$$

$$a^\lambda = \langle [(\mu^-)^\lambda, (\mu^+)^\lambda], \left[\sqrt[q]{1-(1-(\nu^-)^q)^\lambda}, \sqrt[q]{1-(1-(\nu^+)^q)^\lambda}\right] \rangle. \quad (6)$$

**Definition 2.3(Wang *et* al., 2019).** For each IVq-ROFN $a = \langle [\mu^-, \mu^+,], [\nu^-, \nu^+] \rangle$, the score function is defined as:

$$S(a) = \frac{1}{2}[(\mu^-)^q + (\mu^+)^q - (\nu^-)^q - (\nu^+)^q], q \geq 1. \quad (7)$$

**Definition 2.4(Peng and Yong, 2016).** For each IVq-ROFN $a = ([\mu^-, \mu^+,], [\nu^-, \nu^+])$, the accuracy function is defined as

$$H(a) = \frac{1}{2}[(\mu^-)^q + (\mu^+)^q + (\nu^-)^q + (\nu^+)^q], q \geq 1. \quad (8)$$

**Definition 2.5(Peng and Yong, 2016).** Let $a_1 = ([\mu_{a_1}^-, \mu_{a_1}^+], [\nu_{a_1}^-, \nu_{a_1}^+])$ and $a_2 = ([\mu_{a_2}^-, \mu_{a_2}^+], [\nu_{a_2}^-, \nu_{a_2}^+])$ be two IVq-ROFNs. The size comparison rules are defined as:

(1) If $S(a_1) > S(a_2)$, then $a_1 > a_2$;

(2) If $S(a_1) < S(a_2)$, then $a_1 < a_2$;

(3) If $S(a_1) = S(a_2)$, then we further calculate their accuracy and compare the following. ① if $H(a_1) > H(a_2)$, then $a_1 > a_2$; ② if $H(a_1) < H(a_2)$, then $a_1 < a_2$; and ③ if $H(a_1) = H(a_2)$, then $a_1 = a_2$.

### 2.2 Weber operator

**Definition 2.6(Weber, 1983)** For any two real numbers $a, b\epsilon[0,1]$, Weber operator is a t-norm operation, which is defined as

$$T_W(a,b) = \max(\frac{a+b+\lambda ab-1}{1+\lambda}, 0), \lambda\epsilon(-1, +\infty). \quad (9)$$

Its corresponding t-conorm operation is

$$T_W^*(a,b) = \min(a + b + \lambda ab, 1), \lambda\epsilon(-1, +\infty). \quad (10)$$

In Eqs. (9) and (10), $\lambda$ is a variable parameter, which can provide better accuracy, interpretability, and generalization by flexibly adjusting parameters to adapt to different application scenarios and datasets.

## 3 Interval-valued q-Rung Orthopair Fuzzy Weber Operator

This section outlines the IVq-ROFW operations and interprets the IVq-ROFWOWA operator. The Swing-based MAGDM method based on the IVq-ROFWOWA operator is then introduced.

### 3.1 IVq-ROFW operations

Based on the t-norm operation and the t-conorm operation of the Weber operator Eqs. (9) and (10), the IVq-ROFW operations are defined.

**Definition 3.1.** Let $a = ([\mu^-, \mu^+,], [\nu^-, \nu^+])$ , $a_1 = ([\mu_{a_1}^-, \mu_{a_1}^+], [\nu_{a_1}^-, \nu_{a_1}^+])$ and $a_2 = ([\mu_{a_2}^-, \mu_{a_2}^+], [\nu_{a_2}^-, \nu_{a_2}^+])$ be three IVq-ROFNs with $q \geq 1$. the IVq-ROFW operations can be defined as

$$a_1 \oplus_W a_2 = \begin{pmatrix} \begin{bmatrix} \sqrt[q]{min((\mu_{a1}^-)^q + (\mu_{a2}^-)^q + \lambda(\mu_{a1}^-)^q(\mu_{a2}^-)^q, 1)} \\ \sqrt[q]{min((\mu_{a1}^+)^q + (\mu_{a2}^+)^q + \lambda(\mu_{a1}^+)^q(\mu_{a2}^+)^q, 1)} \end{bmatrix}, \\ \begin{bmatrix} \sqrt[q]{max(\frac{(\nu_{a1}^-)^q+(\nu_{a2}^-)^q+\lambda(\nu_{a1}^-)^q(\nu_{a2}^-)^q-1}{1+\lambda}, 0)}, \\ \sqrt[q]{max(\frac{(\nu_{a1}^+)^q+(\nu_{a2}^+)^q+\lambda(\nu_{a1}^+)^q(\nu_{a2}^+)^q-1}{1+\lambda}, 0)} \end{bmatrix} \end{pmatrix}, \quad (11)$$

$$a_1 \otimes_W a_2 = \begin{pmatrix} \begin{bmatrix} \sqrt[q]{max(\frac{(\mu_{a_1}^-)^q + (\mu_{a_2}^-)^q + \lambda(\mu_{a_1}^-)^q(\mu_{a_2}^-)^q - 1}{1 + \lambda}, 0)}, \\ \sqrt[q]{max(\frac{(\mu_{a_1}^+)^q + (\mu_{a_2}^+)^q + \lambda(\mu_{a_1}^+)^q(\mu_{a_2}^+)^q - 1}{1 + \lambda}, 0)} \end{bmatrix}, \\ \begin{bmatrix} \sqrt[q]{min((\nu_{a1}^-)^q + (\nu_{a2}^-)^q + \lambda(\nu_{a1}^-)^q(\nu_{a2}^-)^q, 1)}, \\ \sqrt[q]{min((\nu_{a1}^+)^q + (\nu_{a2}^+)^q + \lambda(\nu_{a1}^+)^q(\nu_{a2}^+)^q, 1)} \end{bmatrix} \end{pmatrix}, \quad (12)$$

$$ka = \left( \begin{bmatrix} \sqrt[q]{min(\frac{1}{\lambda}((1+\lambda(\mu^-)^q)^k - 1),1)}, \\ \sqrt[q]{min(\frac{1}{\lambda}((1+\lambda(\mu^+)^q)^k - 1),1)} \end{bmatrix}, \begin{bmatrix} \sqrt[q]{max(\frac{1}{\lambda}(\frac{(1+\lambda(\nu^-)^q)^k}{(1+\lambda)^{k-1}} - 1),0)}, \\ \sqrt[q]{max(\frac{1}{\lambda}(\frac{(1+\lambda(\nu^+)^q)^k}{(1+\lambda)^{k-1}} - 1),0)}, \end{bmatrix} \right), \tag{13}$$

$$a^k = \left( \begin{bmatrix} \sqrt[q]{max(\frac{1}{\lambda}\left(\frac{(1+\lambda(\mu^-)^q)^k}{(1+\lambda)^{k-1}} - 1\right),0)}, \\ \sqrt[q]{max(\frac{1}{\lambda}\left(\frac{(1+\lambda(\mu^+)^q)^k}{(1+\lambda)^{k-1}} - 1\right),0)} \end{bmatrix}, \begin{bmatrix} \sqrt[q]{min(\frac{1}{\lambda}((1+\lambda(\nu^-)^q)^k - 1),1)} \\ \sqrt[q]{min(\frac{1}{\lambda}((1+\lambda(\nu^+)^q)^k - 1),1)} \end{bmatrix} \right). \tag{14}$$

It can be verified that the IVq-ROFW addition and product in Definition 3.1 satisfy the rules of t-conorms and t-norms. In addition, the multiplication and power satisfy the rules of t-conorms and t-norms.

**Theorem 3.1.** Let $a = ([\mu^-, \mu^+,], [\nu^-, \nu^+])$, $a_1 = ([\mu_{a_1}^-, \mu_{a_1}^+], [\nu_{a_1}^-, \nu_{a_1}^+])$ and $a_2 = ([\mu_{a_2}^-, \mu_{a_2}^+], [\nu_{a_2}^-, \nu_{a_2}^+])$ be three IVq-ROFNs with $q \geq 1$. The IVq-ROFW operations then satisfy the following six properties when $k, k_1, k_2 > 0$.

$$a_1 \oplus_W a_2 = a_2 \oplus_W a_1, \tag{15}$$

$$a_1 \otimes_W a_2 = a_2 \otimes_W a_1, \tag{16}$$

$$k(a_1 \oplus_W a_2) = (ka_1) \oplus_W (ka_2), \tag{17}$$

$$(a_1 \otimes_W a_2)^k = (a_1{}^k) \otimes_W (a_2{}^k), \tag{18}$$

$$(k_1 a) \oplus_W (k_2 a) = (k_1 + k_2)a, \tag{19}$$

$$(a^{k_1}) \otimes_W (a^{k_2}) = a^{(k_1+k_2)}. \tag{20}$$

The proof of Theorem 3.1 can be verified by applying the IVq-ROFW operations. This paper only gives the proof of Eq. (17).

**Proof.** From Eqs. (11) and (14), we have

$(ka_1)\oplus_W(ka_2)$

$$= ka_1\left(\begin{bmatrix}\sqrt[q]{min(\frac{1}{\lambda}((1+\lambda({\mu_{a1}}^{-})^q)^k-1),1)},\\ \sqrt[q]{min(\frac{1}{\lambda}((1+\lambda({\mu_{a1}}^{+})^q)^k-1),1)}\end{bmatrix},\begin{bmatrix}\sqrt[q]{max(\frac{1}{\lambda}(\frac{(1+\lambda({\nu_{a1}}^{-})^q)^k}{(1+\lambda)^{k-1}}-1),0)},\\ \sqrt[q]{max(\frac{1}{\lambda}(\frac{(1+\lambda({\nu_{a1}}^{+})^q)^k}{(1+\lambda)^{k-1}}-1),0)},\end{bmatrix}\right)\oplus_W ka_2\left(\begin{bmatrix}\sqrt[q]{min(\frac{1}{\lambda}((1+\lambda({\mu_{a2}}^{-})^q)^k-1),1)},\\ \sqrt[q]{min(\frac{1}{\lambda}((1+\lambda({\mu_{a2}}^{+})^q)^k-1),1)}\end{bmatrix},\begin{bmatrix}\sqrt[q]{max(\frac{1}{\lambda}(\frac{(1+\lambda({\nu_{a2}}^{-})^q)^k}{(1+\lambda)^{k-1}}-1),0)},\\ \sqrt[q]{max(\frac{1}{\lambda}(\frac{(1+\lambda({\nu_{a2}}^{+})^q)^k}{(1+\lambda)^{k-1}}-1),0)},\end{bmatrix}\right)$$

$$=\left(\begin{array}{l}\begin{bmatrix}\sqrt[q]{min((\sqrt[q]{min(\frac{1}{\lambda}((1+\lambda(\mu_{a1}^{-})^q)^k-1),1)})^q+(\sqrt[q]{min(\frac{1}{\lambda}((1+\lambda(\mu_{a2}^{-})^q)^k-1),1)})^q+\lambda(\sqrt[q]{min(\frac{1}{\lambda}((1+\lambda({\mu_{a1}}^{-})^q)^k-1),1)})^q(\mu_{a2}^{-})^q,1)}\\ \sqrt[q]{min((\sqrt[q]{min(\frac{1}{\lambda}((1+\lambda(\mu_{a1}^{+})^q)^k-1),1)})^q+(\sqrt[q]{min(\frac{1}{\lambda}((1+\lambda(\mu_{a2}^{-})^q)^k-1),1)})^q+\lambda(\sqrt[q]{min(\frac{1}{\lambda}((1+\lambda({\mu_{a1}}^{+})^q)^k-1),1)})^q(\mu_{a2}^{+})^q,1)}\end{bmatrix},\\ \begin{bmatrix}\sqrt[q]{max(\frac{(\sqrt[q]{max(\frac{1}{\lambda}(\frac{(1+\lambda({\nu_{a1}}^{-})^q)^k}{(1+\lambda)^{k-1}}-1),0),})^q+(\sqrt[q]{max(\frac{1}{\lambda}(\frac{(1+\lambda({\nu_{a2}}^{-})^q)^k}{(1+\lambda)^{k-1}}-1),0),})^q+\lambda(\sqrt[q]{max(\frac{1}{\lambda}(\frac{(1+\lambda({\nu_{a1}}^{-})^q)^k}{(1+\lambda)^{k-1}}-1),0),})^q(\sqrt[q]{max(\frac{1}{\lambda}(\frac{(1+\lambda({\nu_{a2}}^{-})^q)^k}{(1+\lambda)^{k-1}}-1),0),})^q-1}{1+\lambda},0)},\\ max(\frac{(\sqrt[q]{max(\frac{1}{\lambda}(\frac{(1+\lambda({\nu_{a1}}^{+})^q)^k}{(1+\lambda)^{k-1}}-1),0),})^q+(\sqrt[q]{max(\frac{1}{\lambda}(\frac{(1+\lambda(\nu_{a2}+)^q)^k}{(1+\lambda)^{k-1}}-1),0),})^q+\lambda(\sqrt[q]{max(\frac{1}{\lambda}(\frac{(1+\lambda({\nu_{a1}}^{+})^q)^k}{(1+\lambda)^{k-1}}-1),0),})^q(\sqrt[q]{max(\frac{1}{\lambda}(\frac{(1+\lambda(\nu_{a2}+)^q)^k}{(1+\lambda)^{k-1}}-1),0),})^q-1}{1+\lambda},0\end{bmatrix}\end{array}\right)$$

$$= k\left(\begin{bmatrix}\sqrt[q]{min((\mu_{a1}^{-})^q+(\mu_{a2}^{-})^q+\lambda(\mu_{a1}^{-})^q(\mu_{a2}^{-})^q,1)}\\ \sqrt[q]{min((\mu_{a1}^{+})^q+(\mu_{a2}^{+})^q+\lambda(\mu_{a1}^{+})^q(\mu_{a2}^{+})^q,1)}\end{bmatrix},\begin{bmatrix}\sqrt[q]{max(\frac{(v_{a_1}^{-})^q+(v_{a_2}^{-})^q+\lambda(v_{a_1}^{-})^q(v_{a_2}^{-})^q-1}{1+\lambda},0)},\\ \sqrt[q]{max(\frac{(v_{a_1}^{+})^q+(v_{a_2}^{+})^q+\lambda(v_{a_1}^{+})^q(v_{a_2}^{+})^q-1}{1+\lambda},0)}\end{bmatrix}\right)=k(a_1\oplus_W a_2).$$

## 3.2 IVq-ROFWOWA operator

**Definition 3.2.** Let $a_i=([\mu_{a_i}^{-},\mu_{a_i}^{+}],[v_{a_i}^{-},v_{a_i}^{+}])(i=1,2,3,\cdots,n)$ be a group of IVq-ROFNs, and $\omega=(\omega_1,\omega_2,\omega_3,\cdots,\omega_n)^T$ be the weight vector, which satisfy$\sum_{i=1}^{n}\omega_i=1,\omega_i\geq 0(i=1,2,3,\cdots,n)$. The IVq-ROFWOWA operator shown in Eq. (21) is a mapping from $\mathrm{IVq}-\mathrm{ROFNs}^n$ to $\mathrm{IVq}-\mathrm{ROFN}$:

$$IVq-ROFWOWA(a_1,a_2,a_3,\cdots,a_n)=\underset{i=1}{\overset{n}{\oplus_W}}\omega_i a_{\sigma(i)}, \tag{21}$$

where $(\sigma(1),\sigma(2),\sigma(3),\cdots,\sigma(n))$ is a replacement of $(1,2,3,\cdots,n)$, which satisfies $a_{\sigma(i)}\geq a_{\sigma(i+1)}(i=1,2,3,\cdots,n)$.

**Theorem 3.2.** Let $a_i=\langle[\mu_{a_i}^{-},\mu_{a_i}^{+}],[v_{a_i}^{-},v_{a_i}^{+}]\rangle(i=1,2,3,\cdots,n)$ be a group of IVq-ROFNs. The IVq-ROFWOWA operator remains an IVq-ROFN and is shown in Eq. (22):

$IVq-ROFWOWA(a_1,a_2,a_3,\cdots,a_n)$

$$=\left(\begin{bmatrix}\sqrt[q]{min(\frac{1}{\lambda}\left(\prod_{i=1}^{n}\left(1+\lambda(\mu_{\vartheta(i)}^{-})^q\right)-1\right),1)},\\ \sqrt[q]{min(\frac{1}{\lambda}\left(\prod_{i=1}^{n}\left(1+\lambda(\mu_{\vartheta(i)}^{+})^q\right)-1\right),1)}\end{bmatrix},\begin{bmatrix}\sqrt[q]{max(\frac{1}{\lambda}(\frac{\prod_{i=1}^{n}\left(1+\lambda(v_{\vartheta(i)}^{-})^q\right)}{(1+\lambda)^{n-1}}-1),0)},\\ \sqrt[q]{max(\frac{1}{\lambda}(\frac{\prod_{i=1}^{n}\left(1+\lambda(v_{\vartheta(i)}^{+})^q\right)}{(1+\lambda)^{n-1}}-1),0)},\end{bmatrix}\right), \tag{22}$$

where $\left(\sigma(1),\sigma(2),\sigma(3),\cdots,\sigma(n)\right)$ satisfying $a_{\sigma(i)} \geq a_{\sigma(i+1)} (i = 1,2,3,\cdots,n)$. Additionally, $a_{\sigma(i)} = [\mu_{\sigma(i)}^{-}, \mu_{\sigma(i)}^{+}], [v_{\sigma(i)}^{-}, v_{\sigma(i)}^{+}]$. Meanwhile, we have following definitions to simplify Eq. (22).

$$\mu_{\theta(i)}^{-} = \sqrt[q]{min(\frac{1}{\lambda}((1+\lambda(\mu_{a(i)}^{-})^{q})^{w_i}-1),1)},\quad \mu_{\theta(i)}^{+} = \sqrt[q]{min(\frac{1}{\lambda}((1+\lambda(\mu_{a(i)}^{+})^{q})^{w_i}-1),1)},$$

$$v_{\theta(i)}^{-} = \sqrt[q]{max(\frac{1}{\lambda}(\frac{(1+\lambda(v_{a(i)}^{-})^{q})^{w_i}}{(1+\lambda)^{w_i-1}}-1),0)},\qquad v_{\theta(i)}^{+} = \sqrt[q]{max(\frac{1}{\lambda}(\frac{(1+\lambda(v_{a(i)}^{+})^{q})^{w_i}}{(1+\lambda)^{w_i-1}}-1),0)}.$$

**Proof** Theorem 3.2 can be proved by mathematical induction.

Assume that $a_i \geq a_{i+1} (i = 1,2,3,\cdots,n-1)$, i.e., $\left(\sigma(1),\sigma(2),\sigma(3),\cdots,\sigma(n)\right) = (1,2,3,\cdots,n)$. We consider three steps.

① For $n = 2$, we have

$$w_1 a_1 = \left(\begin{bmatrix}\sqrt[q]{min(\frac{1}{\lambda}((1+\lambda(\mu_{a_1}{}^{-})^{q})^{w_1}-1),1)},\\ \sqrt[q]{min(\frac{1}{\lambda}((1+\lambda(\mu_{a_1}{}^{+})^{q})^{w_1}-1),1)}\end{bmatrix}, \begin{bmatrix}\sqrt[q]{max(\frac{1}{\lambda}(\frac{(1+\lambda(v_{a_1}{}^{-})^{q})^{W_1}}{(1+\lambda)^{k-1}}-1),0)},\\ \sqrt[q]{max(\frac{1}{\lambda}(\frac{(1+\lambda(v_{a_1}{}^{+})^{q})^{W_1}}{(1+\lambda)^{k-1}}-1),0)},\end{bmatrix}\right)$$

$$w_2 a_2 = \left(\begin{bmatrix}\sqrt[q]{min(\frac{1}{\lambda}((1+\lambda(\mu_{a_2}{}^{-})^{q})^{w_2}-1),1)},\\ \sqrt[q]{min(\frac{1}{\lambda}((1+\lambda(\mu_{a_2}{}^{+})^{q})^{w_2}-1),1)}\end{bmatrix}, \begin{bmatrix}\sqrt[q]{max(\frac{1}{\lambda}(\frac{(1+\lambda(v_{a_2}{}^{-})^{q})^{W_2}}{(1+\lambda)^{k-1}}-1),0)},\\ \sqrt[q]{max(\frac{1}{\lambda}(\frac{(1+\lambda(v_{a_2}{}^{+})^{q})^{W_2}}{(1+\lambda)^{k-1}}-1),0)},\end{bmatrix}\right).$$

Then we can get

$$IVq-ROFWOWA(a_1,a_2) = \overset{2}{\underset{i=1}{\oplus_W}} \omega_i a_i$$

$$= \left(\begin{bmatrix}\sqrt[q]{min((\sqrt[q]{min(\frac{1}{\lambda}((1+\lambda(\mu_{a1}^{-})^{q})^{w_1}-1),1)})^{q}+(\sqrt[q]{min(\frac{1}{\lambda}((1+\lambda(\mu_{a2}^{-})^{q})^{w_2}-1),1)})^{q}+\lambda(\sqrt[q]{min(\frac{1}{\lambda}((1+\lambda(\mu_{a1}{}^{-})^{q})^{w_1}-1),1)})^{q}(\sqrt[q]{min(\frac{1}{\lambda}((1+\lambda(\mu_{a2}^{-})^{q})^{w_2}-1),1)})^{q},1)}\\ \sqrt[q]{min((\sqrt[q]{min(\frac{1}{\lambda}((1+\lambda(\mu_{a1}^{+})^{q})^{w_1}-1),1)})^{q}+(\sqrt[q]{min(\frac{1}{\lambda}((1+\lambda(\mu_{a2}^{-})^{q})^{w_2}-1),1)})^{q}+\lambda(\sqrt[q]{min(\frac{1}{\lambda}((1+\lambda(\mu_{a1}{}^{+})^{q})^{w_1}-1),1)})^{q}(\sqrt[q]{min(\frac{1}{\lambda}((1+\lambda(\mu_{a2}^{+})^{q})^{w_2}-1),1)})^{q},1)}\end{bmatrix}, \begin{bmatrix}\sqrt[q]{max(\frac{(\sqrt[q]{max(\frac{1}{\lambda}(\frac{(1+\lambda(v_{a1}{}^{-})^{q})^{w_1}}{(1+\lambda)^{k-1}}-1),0),})^{q}+(\sqrt[q]{max(\frac{1}{\lambda}(\frac{(1+\lambda(v_{a2}{}^{-})^{q})^{w_2}}{(1+\lambda)^{k-1}}-1),0),})^{q}+\lambda(\sqrt[q]{max(\frac{1}{\lambda}(\frac{(1+\lambda(v_{a1}{}^{-})^{q})^{w_1}}{(1+\lambda)^{k-1}}-1),0),})^{q}(\sqrt[q]{max(\frac{1}{\lambda}(\frac{(1+\lambda(v_{a2}{}^{-})^{q})^{w_2}}{(1+\lambda)^{k-1}}-1),0),})^{q}-1}{1+\lambda},0)},\\ \sqrt[q]{max(\frac{(\sqrt[q]{max(\frac{1}{\lambda}(\frac{(1+\lambda(v_{a1}{}^{+})^{q})^{w_1}}{(1+\lambda)^{k-1}}-1),0),})^{q}+(\sqrt[q]{max(\frac{1}{\lambda}(\frac{(1+\lambda(v_{a2}{}^{+})^{q})^{w_2}}{(1+\lambda)^{k-1}}-1),0),})^{q}+\lambda(\sqrt[q]{max(\frac{1}{\lambda}(\frac{(1+\lambda(v_{a1}{}^{+})^{q})^{w_1}}{(1+\lambda)^{k-1}}-1),0),})^{q}(\sqrt[q]{max(\frac{1}{\lambda}(\frac{(1+\lambda(v_{a2}{}^{+})^{q})^{w_2}}{(1+\lambda)^{k-1}}-1),0),})^{q}-1}{1+\lambda},0}\end{bmatrix}\right)$$

$$= \left(\begin{bmatrix}\sqrt[q]{min(\frac{1}{\lambda}\left(\prod_{i=1}^{n}\left(1+\lambda(\mu_{\vartheta_{(i)}}^{-})^{q}\right)-1\right),1)},\\ \sqrt[q]{min(\frac{1}{\lambda}\left(\prod_{i=1}^{n}\left(1+\lambda(\mu_{\vartheta_{(i)}}^{+})^{q}\right)-1\right),1)}\end{bmatrix}, \begin{bmatrix}\sqrt[q]{max(\frac{1}{\lambda}(\frac{\prod_{i=1}^{n}\left(1+\lambda(v_{\vartheta_{(i)}}^{-})^{q}\right)^{\square}}{(1+\lambda)^{n-1}}-1),0)},\\ \sqrt[q]{max(\frac{1}{\lambda}(\frac{\prod_{i=1}^{n}\left(1+\lambda(v_{\vartheta_{(i)}}^{+})^{q}\right)^{\square}}{(1+\lambda)^{n-1}}-s1),0)},\end{bmatrix}\right).$$

Among them

$$\mu_{\theta(i)}^{-} = \sqrt[q]{min(\frac{1}{\lambda}((1+\lambda(\mu_{a(i)}^{-})^{q})^{w_i}-1),1)},\quad \mu_{\theta(i)}^{+} = \sqrt[q]{min(\frac{1}{\lambda}((1+\lambda(\mu_{a(i)}^{+})^{q})^{w_i}-1),1)},$$

$$v_{\theta(i)}^{-}=\sqrt[q]{max(\frac{1}{\lambda}(\frac{(1+\lambda(v_{a(i)}^{-})^q)^{w_i}}{(1+\lambda)^{w_i-1}}-1),0)},\ v_{\theta(i)}^{+}=\sqrt[q]{max(\frac{1}{\lambda}(\frac{(1+\lambda(v_{a(i)}^{+})^q)^{w_i}}{(1+\lambda)^{w_i-1}}-1),0)}.$$

This implies that Eq. (22) holds for $n$=2.

② Assuming that Eq. (22) holds for $n=k$, we can get $IVq-ROFWOWA(a_1,a_2,a_3,\cdots,a_k)$ as follows:

$IVq-ROFWOWA(a_1,a_2,a_3,\cdots,a_k)=$

$$\left(\begin{bmatrix}\sqrt[q]{min(\frac{1}{\lambda}\left(\prod_{i=1}^{n}\left(1+\lambda(\mu_{\vartheta_{(i)}}^{-})^q\right)-1\right),1)},\\ \sqrt[q]{min(\frac{1}{\lambda}\left(\prod_{i=1}^{n}\left(1+\lambda(\mu_{\vartheta_{(i)}}^{+})^q\right)-1\right),1)}\end{bmatrix},\begin{bmatrix}\sqrt[q]{max(\frac{1}{\lambda}(\frac{\prod_{i=1}^{n}\left(1+\lambda(v_{\vartheta_{(i)}}^{-})^q\right)}{(1+\lambda)^{n-1}}-1),0)},\\ \sqrt[q]{max(\frac{1}{\lambda}(\frac{\prod_{i=1}^{n}\left(1+\lambda(v_{\vartheta_{(i)}}^{+})^q\right)}{(1+\lambda)^{n-1}}-1),0)},\end{bmatrix}\right)$$

Among them

$$\mu_{\theta(i)}^{-}=\sqrt[q]{min(\frac{1}{\lambda}((1+\lambda(\mu_{a(i)}^{-})^q)^{w_i}-1),1)},\ \mu_{\theta(i)}^{+}=\sqrt[q]{min(\frac{1}{\lambda}((1+\lambda(\mu_{a(i)}^{+})^q)^{w_i}-1),1)},$$

$$v_{\theta(i)}^{-}=\sqrt[q]{max(\frac{1}{\lambda}(\frac{(1+\lambda(v_{a(i)}^{-})^q)^{w_i}}{(1+\lambda)^{w_i-1}}-1),0)},v_{\theta(i)}^{+}=\sqrt[q]{max(\frac{1}{\lambda}(\frac{(1+\lambda(v_{a(i)}^{+})^q)^{w_i}}{(1+\lambda)^{w_i-1}}-1),0)}.$$

③ When $n=k+1$, we have $IVq-ROFWOWA(a_1,a_2,a_3,\cdots,a_k,a_{k+1})$ as follows:

$IVq-ROFWOWA(a_1,a_2,a_3,\cdots,a_k,a_{k+1})$

$=IVq-ROFWOWA(a_1,a_2,a_3,\cdots,a_k)\oplus_W\omega_{k+1}a_{k+1}$

$$=\left(\begin{bmatrix}\sqrt[q]{min(\frac{1}{\lambda}\left(\prod_{i=1}^{n}\left(1+\lambda(\mu_{\vartheta_{(i)}}^{-})^q\right)-1\right),1)},\\ \sqrt[q]{min(\frac{1}{\lambda}\left(\prod_{i=1}^{n}\left(1+\lambda(\mu_{\vartheta_{(i)}}^{+})^q\right)-1\right),1)}\end{bmatrix},\begin{bmatrix}\sqrt[q]{max(\frac{1}{\lambda}\left(\frac{\prod_{i=1}^{n}\left(1+\lambda(v_{\vartheta_{(i)}}^{-})^q\right)}{(1+\lambda)^{n-1}}-1\right),0)},\\ \sqrt[q]{max(\frac{1}{\lambda}\left(\frac{\prod_{i=1}^{n}\left(1+\lambda(v_{\vartheta_{(i)}}^{+})^q\right)}{(1+\lambda)^{n-1}}-1\right),0)},\end{bmatrix}\right)\oplus_W$$

$$\left(\begin{bmatrix}\sqrt[q]{min(\frac{1}{\lambda}((1+\lambda(\mu_{\theta(k+1)}^{-})^q)^{w_{i+1}}-1),1)},\\ \sqrt[q]{min(\frac{1}{\lambda}((1+\lambda(\mu_{\theta(k+1)}^{+})^q)^{w_{i+1}}-1),1)}\end{bmatrix},\begin{bmatrix}\sqrt[q]{max(\frac{1}{\lambda}(\frac{(1+\lambda(v^{-})^q)^{w_{i+1}}}{(1+\lambda)^{k-1}}-1),0)},\\ \sqrt[q]{max(\frac{1}{\lambda}(\frac{(1+\lambda(v^{+})^q)^{w_{i+1}}}{(1+\lambda)^{k-1}}-1),0)},\end{bmatrix}\right)$$

$$=\left(\begin{bmatrix}\sqrt[q]{\min(\frac{1}{\lambda}\left(\prod_{i=1}^{n+1}\left(1+\lambda(\mu_{\theta_{(i)}}^{-})^q\right)-1\right),1)},\\ \sqrt[q]{\min(\frac{1}{\lambda}\left(\prod_{i=1}^{n+1}\left(1+\lambda(\mu_{\theta_{(i)}}^{+})^q\right)-1\right),1)}\end{bmatrix},\begin{bmatrix}\sqrt[q]{\max(\frac{1}{\lambda}(\frac{\prod_{i=1}^{n+1}\left(1+\lambda(v_{\theta_{(i)}}^{-})^q\right)}{(1+\lambda)^{n-1}}-1),0)},\\ \sqrt[q]{\max(\frac{1}{\lambda}(\frac{\prod_{i=1}^{n+1}\left(1+\lambda(v_{\theta_{(i)}}^{+})^q\right)}{(1+\lambda)^{n-1}}-1),0)},\end{bmatrix}\right)$$

Among them

$\mu_{\theta(i)}^{-} = \sqrt[q]{min(\frac{1}{\lambda}((1+\lambda(\mu_{a(i)}^{-})^{q})^{w_i}-1),1)}$, $\mu_{\theta(i)}^{+} = \sqrt[q]{min(\frac{1}{\lambda}((1+\lambda(\mu_{a(i)}^{+})^{q})^{w_i}-1),1)}$,

$v_{\theta(i)}^{-} = \sqrt[q]{max(\frac{1}{\lambda}(\frac{(1+\lambda(v_{a(i)}^{-})^{q})^{w_i}}{(1+\lambda)^{w_i-1}}-1),0)}$, $v_{\theta(i)}^{+} = \sqrt[q]{max(\frac{1}{\lambda}(\frac{(1+\lambda(v_{a(i)}^{+})^{q})^{w_i}}{(1+\lambda)^{w_i-1}}-1),0)}$.

That is, Eq. (22) holds for $n = k + 1$. Thus, by the principle of mathematical induction, Theorem 3.2 can be verified.

It can be proved that the IVq-ROFWOWA operator satisfies idempotency, commutativity, monotonicity, and boundedness, which are introduced in detail in Theorems 3.3-3.6. Because it is not difficult to prove these properties, the proof of them is omitted in this paper by using the IVq-ROFW operations in Eqs. (11) ~ (14).

**Theorem 3.3 (Idempotency).** Let $a_i = ([\mu_{a_i}^{-}, \mu_{a_i}^{+}], [v_{a_i}^{-}, v_{a_i}^{+}]), (i = 1,2,3,\cdots,n)$ be a group of IVq-ROFNs. The IVq-ROFWOWA operator then satisfies the idempotency, i.e., Eq. (23) holds when $a = a_1 = a_2 = \cdots = a_n$ and their attribute weight is $\omega = (\omega_1, \omega_2, \omega_3, \cdots, \omega_n)$ satisfying $\sum_{j=1}^{n} \omega_j = 1, \omega_j \epsilon [0,1], (j = 1,2,3,\cdots,n)$:

$$IVq-ROFWOWA(a_1, a_2, a_3, \cdots, a_n) = a. \quad (23)$$

**Proof.** Theorem 3.3 can be proved by IVq-ROFWOWA that satisfies the rules of t-conorms and t-norms:

$IVq-ROFWOWA(a_1, a_2, a_3, \cdots, a_n)$

$=< \left[g^{-1}\left(w_1 g(\mu_{a_1}^{-}) + w_2 g(\mu_{a_2}^{-}) + \ldots\ldots + w_n g(\mu_{a_n}^{-})\right), g^{-1}\left(w_1 g(\mu_{a_1}^{+}) + w_2 g(\mu_{a_2}^{+}) + \ldots\ldots + w_n g(\mu_{a_n}^{+})\right)\right],$

$\left[f^{-1}\left(w_1 f(v_{a_1}^{-}) + w_2 f(v_{a_2}^{-}) + \ldots\ldots + w_n f(v_{a_n}^{-})\right), f^{-1}\left(w_1 f(v_{a_1}^{+}) + w_2 f(v_{a_2}^{+}) + \ldots\ldots + w_n f(v_{a_n}^{+})\right)\right] >$

$=\langle[\mu_{a}^{-}, \mu_{a}^{+}], [v_{a}^{-}, v_{a}^{+}]\rangle.$

**Theorem 3.4 (Commutativity).** Let $a_i = \langle[\mu_{a_i}^{-}, \mu_{a_i}^{+}], [v_{a_i}^{-}, v_{a_i}^{+}]\rangle, (i = 1,2,3,\cdots,n)$ be a group of IVq-ROFNs. The IVq-ROFWOWA operator then satisfies the commutativity, i.e., Eq. (24) holds when $(a_1', a_2', a_3', \cdots, a_n')$ is an arbitrary replacement of $(a_1, a_2, a_3, \cdots, a_n)$:

$$IVq-ROFWOWA(a_1', a_2', a_3', \cdots, a_n') = IVq-ROFWOWA(a_1, a_2, a_3, \cdots, a_n). \quad (24)$$

**Theorem 3.5 (Monotonicity).** Let $a_i = \langle[\mu_{a_i}^{-}, \mu_{a_i}^{+}], [v_{a_i}^{-}, v_{a_i}^{+}]\rangle, (i = 1,2,3,\cdots,n)$ and $b_i = \langle[\mu_{b_i}^{-}, \mu_{b_i}^{+}], [v_{b_i}^{-}, v_{b_i}^{+}]\rangle, (i = 1,2,3,\cdots,n)$ be two groups of IVq-ROFNs. The IVq-ROFWOWA operator then satisfies the monotonicity, i.e., Eq. (25) holds if $\mu_{a_i}^- \leq \mu_{b_i}^-$, $\mu_{a_i}^- \leq \mu_{b_i}^-$, $\nu_{a_i}^- \geq \nu_{b_i}^-$, $\nu_{a_i}^+ \geq \nu_{b_i}^+ (i = 1,2,3,\cdots,n)$:

$$IVq-ROFWOWA(a_1,a_2,a_3,\cdots,a_n) \leq IVq-ROFWOWA(b_1,b_2,b_3,\cdots,b_n). \quad (25)$$

It is worth noting that the monotonicity given here needs to satisfy the size comparison between the IVq-ROFNs. When the data does not satisfy this relationship, the monotonicity does not necessarily hold.

**Theorem 3.6 (Boundedness).** Let $a_i = \langle [\mu_{a_i}^-, \mu_{a_i}^+], [\nu_{a_i}^-, \nu_{a_i}^+] \rangle, (i = 1,2,3,\cdots,n)$ be a group of IVq-ROFNs. The IVq-ROFWOWA operator then satisfies the boundedness, i.e., Eq. (26) holds for $a_{min} = \left(\left[\min_i\{\mu_{a_i}^-\}, \min_i\{\mu_{a_i}^+\}\right], \left[\min_i\{\nu_{a_i}^-\}, \min_i\{\nu_{a_i}^+\}\right]\right)$ and $a_{max} = (\left[\max_i\{\mu_{a_i}^-\}, \max_i\{\mu_{a_i}^+\}\right], \left[\max_i\{\nu_{a_i}^-\}, \max_i\{\nu_{a_i}^+\}\right])$:

$$a_{min} \leq IVq-ROFIDPOWA(a_1,a_2,a_3,\cdots,a_n) \leq a_{max}. \quad (26)$$

### 3.3 Swing-based MAGDM method Based on the IVq-ROFWOWA Operator

This section will propose the group decision-making method based on the IVq-ROFWOWA operator and collaborative filtering of Swing, which can recommend similarities and differences within groups and assign different individuals different attribute weights, thus achieving personalized group decision-making. For a MAGDM problem, let $X = \{x_1,x_2,x_3,\cdots,x_m\}$ be the alternative set and $C = \{c_1,c_2,c_3,\cdots,c_n\}$ be the attribute set whose attribute weight is $\omega = (\omega_1,\omega_2,\omega_3,\cdots,\omega_n)$ satisfying $\sum_{j=1}^{n}\omega_j = 1, \omega_j \epsilon [0,1], (j = 1,2,3,\cdots,n)$. $D = \{d_1,d_2,d_3,\cdots,d_t\}$ is the expert set whose expert weight is $\varphi = (\varphi_{1,}\varphi_{2,}\ldots\ldots\varphi_t)$ satisfying. $\sum_{k=1}^{t}\varphi_k = 1, \varphi_k \epsilon [0,1], (k = 1,2,3,\cdots,t)$. The decision matrix of the $k$-th expert is $A^{(k)} = \left(a_{ij}^{(k)}\right)_{m\times n}$ given by Eq. (35), in which the element $a_{ij}^{(k)}$ represents the decision value of the $k$-th expert on the attribute $j$ of the alternative $i$. $a_{ij}^{(k)}$ is an IVq-ROFN that satisfies $\left(\mu_{a_{ij}^{(k)}}^+\right)^q + \left(\nu_{a_{ij}^{(k)}}^+\right)^q \leq 1, q \geq 1$. It should be noted that the value of $q$ is adapted to the actual situation.

$$A^{(k)} = \begin{bmatrix} a_{11}^{(k)} \cdots a_{1j}^{(k)} \cdots a_{1n}^{(k)} \\ \ddots \quad \ddots \ \vdots \\ a_{ij}^{(k)} \cdots a_{ij}^{(k)} \cdots a_{in}^{(k)} \\ \vdots \ \ddots \ \vdots \ \ddots \ \vdots \\ a_{m1}^{(k)} \cdots a_{mj}^{(k)} \cdots a_{mn}^{(k)} \end{bmatrix}_{m\times n}. \tag{27}$$

### 3.3.1 Attribute weights based on Swing

The advantages of using collaborative filtering to determine attribute weights in MAGDM mainly lie in its ability to facilitate information sharing, build consensus, calculate personalized weights, and discover decision trends. The following steps explain how it works.

(1) According to the expert weights $\varphi = (\varphi_{1,}\varphi_{2,}\ldots\ldots\varphi_t)$ given in Eq. (22). The set $A^{(k)}$ obtains $\mathrm{R} = (\mathrm{r_{ij}})_{\mathrm{m\times n}}$ in Eq. (28).

$$R = \begin{bmatrix} r_{11} \cdots r_{1j} \cdots r_{1n} \\ \ddots \quad \ddots \ \vdots \\ r_{ij} \cdots r_{ij} \cdots r_{in} \\ \vdots \ \ddots \ \vdots \ \ddots \ \vdots \\ r_{m1} \cdots r_{mj} \cdots r_{mn} \end{bmatrix}_{m\times n}. \tag{28}$$

(2) According to Eq. (29), the distance of each element in $R = (r_{ij})_{m\times n}$ to the positive ideal point PIS=([1,1],[0,0]) is calculated as $d_{ij}$ in Eq. (30). $d_{ij}$ can be used to construct a set $D = (d_{ij})_{m\times n}$ in Eq. (31).

$$d(a_1,a_2) = \frac{1}{4}(|(u_{a_1}^-)^q - (u_{a_2}^-)^q| + |(u_{a_1}^+)^q - (u_{a_2}^+)^q| + |(v_{a_1}^-)^q - (v_{a_2}^-)^q| + |(\pi_{a_1}^-)^q - (\pi_{a_2}^-)^q| + |(\pi_{a_1}^+)^q - (\pi_{a_2}^+)^q|, \tag{29}$$

$$\mathrm{d_{ij}} = \mathrm{d}(\mathrm{r_{ij}}, \mathrm{PIS}), \tag{30}$$

$$D = \begin{bmatrix} d_{11} \cdots d_{1j} \cdots d_{1n} \\ \ddots \quad \ddots \ \vdots \\ d_{i1} \cdots d_{ij} \cdots d_{in} \\ \vdots \ \ddots \ \vdots \ \ddots \ \vdots \\ d_{m1} \cdots d_{mj} \cdots d_{mn} \end{bmatrix}_{m\times n}. \tag{31}$$

(3) A threshold value, $d_{bound}$, stands for that they are connected. $b_{ij}$ in Eq. (32) belongs to the Boolean type and their collection matrix $B$ is given in Eq. (33).

$$b_{ij} = \begin{cases} 0, d_{bound} - d_{ij} > 0 \\ 1, d_{bound} - d_{ij} < 0 \end{cases}, \tag{32}$$

$$B = \begin{bmatrix} b_{11} \cdots b_{1j} \cdots b_{1n} \\ \ddots \quad \ddots \quad \vdots \\ b_{i1} \cdots b_{ij} \cdots b_{in} \\ \vdots \ \ddots \ \vdots \ \ddots \ \vdots \\ b_{m1} \cdots b_{mj} \cdots b_{mn} \end{bmatrix}_{m \times n} . \tag{33}$$

(4) By utilizing Eq. (33) to builds swing, we can further obtain $Ts_{ij}$.

(5) Importance between each attribute can be obtained in Eq. (34), where $U_{c_i}$ denotes the set of selected factors $c_i$ alternative, $\varepsilon$ and $\gamma$ are both elements of $U_{c_i} \cap U_{c_j}$, $I_\varepsilon$ and $I_\gamma$ correspond to the set of selected factors $\varepsilon$ and $\gamma$, respectively, and α is the smoothing factor. The attribute relative degree matrix Ts is represented in Eq. (35) and its structure is shown in **Figure 2**, where the lines represent the selection of attributes for each solution with thicker lines indicating a higher degree of selection.

$$Ts_{ij} = \begin{cases} \sum_{\varepsilon \in U_{c_i} \cap U_{c_j}} \sum_{\gamma \in U_{c_i} \cap U_{c_j}} \frac{1}{\sqrt{I_\varepsilon}} \times \frac{1}{\sqrt{I_\gamma}} \frac{1}{\alpha + |I_\varepsilon \cap I_\gamma|} & i! = j \\ 1 & \text{i=}j \end{cases}, \tag{34}$$

$$Ts = \begin{bmatrix} Ts_{11} \cdots Ts_{1j} \cdots Ts_{1n} \\ \ddots \quad \ddots \quad \vdots \\ Ts_{i1} \cdots Ts_{ij} \cdots Ts_{in} \\ \vdots \ \ddots \ \vdots \ \ddots \ \vdots \\ Ts_{m1} \cdots Ts_{mj} \cdots Ts_{mn} \end{bmatrix}_{m \times n} . \tag{35}$$

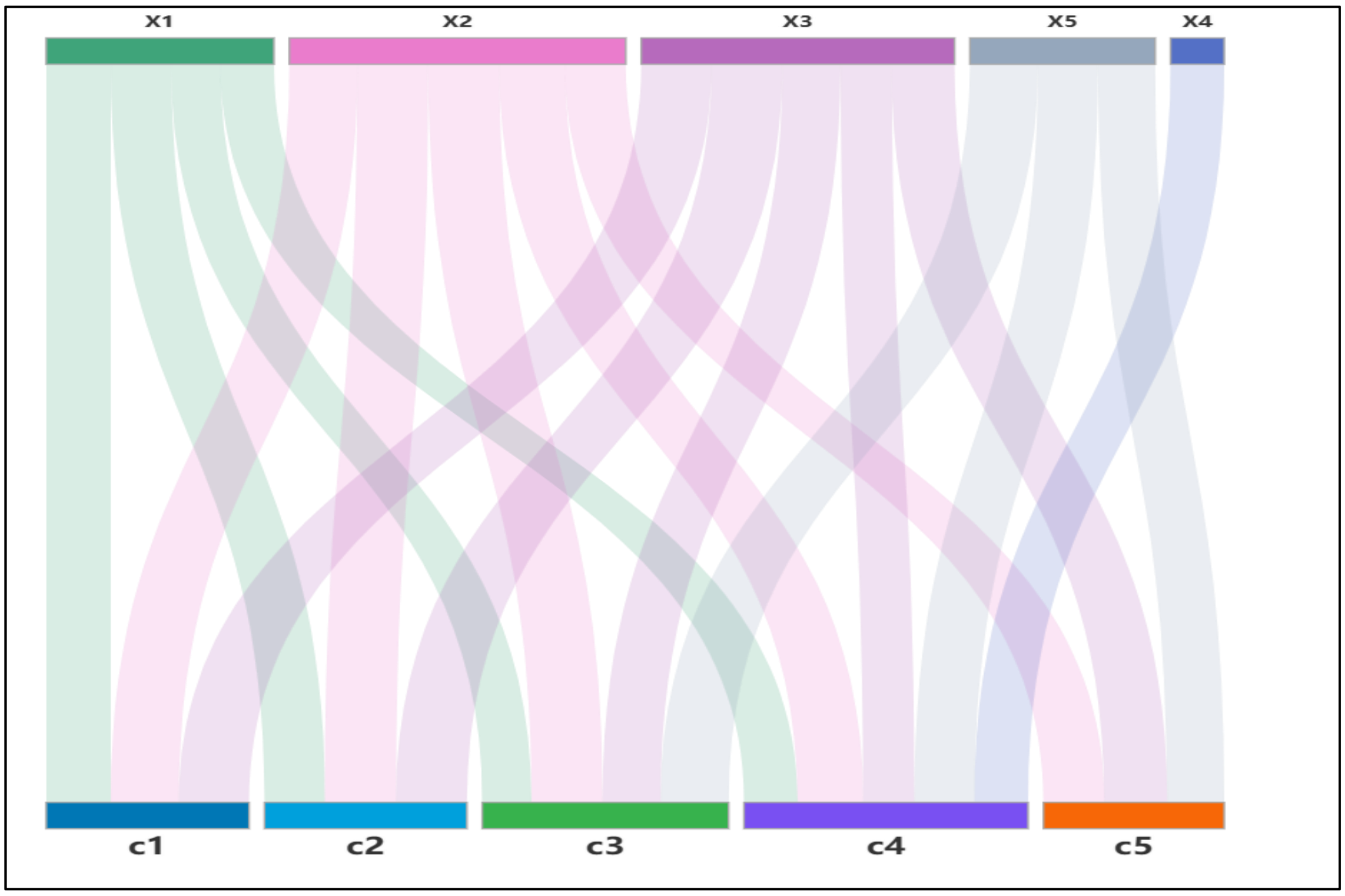

**Figure 2** Connected Matrix

(6) The importance $S_i$ is calculated by Eq. (36):

$$S_i = \frac{1}{n}\sum_{j=1}^{j=n} s_{ij}. \tag{36}$$

(7) Attribute weights vector $w_{Swing} = (w_1, w_2, \ldots, w_n)$ can be obtained by Eq. (37).

$$w_i = \frac{S_i}{\sum_{i=1}^{i=n} S_i}. \tag{37}$$

### 3.3.2 MAGDM based on IVq-ROFWOWA operator

Specific steps of the group decision-making processing method are given in the following steps and **Figure 3**.

(1) Convert expert judgment value to IVq-ROFNs.

In addition to make it easier for experts to make reasonable evaluations in group decision-making, grades can be used for scoring, saving experts from understanding fuzzy theory and reducing the evaluation cost. Inspired by literature [32-33], the evaluation grade is divided into ten scales: certainly low important (CLI), very low important (VLI), lower important (LI), below average important (BAI), average important (AI), above average important (AAI), high important (HI), very high important (VHI), certainly high important (CHI) and exactly equal (EE). It is then transformed into IVq-ROFNs according to the evaluation level as shown in **Table 1**.

**Table 1** Interval-valued q-Rung Orthopair Fuzzy Number Transformations

| Linguistic term | Interval-valued q-rung orthopair fuzzy numbers | | | |
|---|---|---|---|---|
| | $\mu_L$ | $\mu_U$ | $v_L$ | $v_U$ |
| Certainly low important (CLI) | 0.05 | 0.05 | 0.90 | 0.95 |
| Very low important (VLI) | 0.10 | 0.20 | 0.80 | 0.90 |
| Low important (LI) | 0.20 | 0.35 | 0.65 | 0.80 |
| Below average important (BAI) | 0.35 | 0.45 | 0.55 | 0.65 |
| Average important (AI) | 0.45 | 0.55 | 0.45 | 0.55 |
| Above average important (AAI) | 0.55 | 0.65 | 0.35 | 0.45 |
| High important (HI) | 0.65 | 0.80 | 0.20 | 0.35 |
| Very high important (VHI) | 0.80 | 0.90 | 0.10 | 0.20 |
| Certainly high important (CHI) | 0.90 | 0.95 | 0.05 | 0.05 |
| Exactly equal (EE) | 0.1965 | 0.1965 | 0.1965 | 0.1965 |

(2) Select $q$ value. A suitable $q$ is chosen so that all elements in $A^{(k)}(k = 1,2,3,\cdots,t)$ satisfy $\left(\mu^{+}_{a_{ij}^{(k)}}\right)^q + \left(v^{+}_{a_{ij}^{(k)}}\right)^q \leq 1(q \geq 1)$. The value of $q$ that meets the requirements can be found by traversal method.

(3) Aggregate expert judgement value. The standard evaluation matrix $A^{(k)}$ is aggregated by using the IVq-ROFWOWA operator combined with the expert weight $\varphi$. The aggregated matrix $R = \left(r_{ij}\right)_{m\times n}$ is then obtained. The aggregation process is shown in Eq. (38)

$$r_{ij} = IVq - ROFWOWA\left(a_{ij}^{(1)}, a_{ij}^{(2)}, a_{ij}^{(3)}, \cdots, a_{ij}^{(t)}\right). \quad (38)$$

(4) Get aggregation value of $r_i$. Attribute weights are derived by the Swing method by equation (37), the aggregated matrix $R$ is obtained by using the IVq-ROFWOWA operator combined with the attribute weight $\omega$. The aggregation process is shown in Eq. (39):

$$r_i = IVq - ROFWOWA(r_{i1}, r_{i2}, r_{i3}, \cdots, r_{in}). \quad (39)$$

(5) Calculate the score and accuracy value of $r_i$ by Eqs. (7) and (8). $r_i$ is then sorted according to definition 2.5.

(6) Based on the sorting result of $r_i$ in step (4), the sorting result of the alternative $x_i$ is obtained. The most prominent alternative is then selected as the optimal alternative.

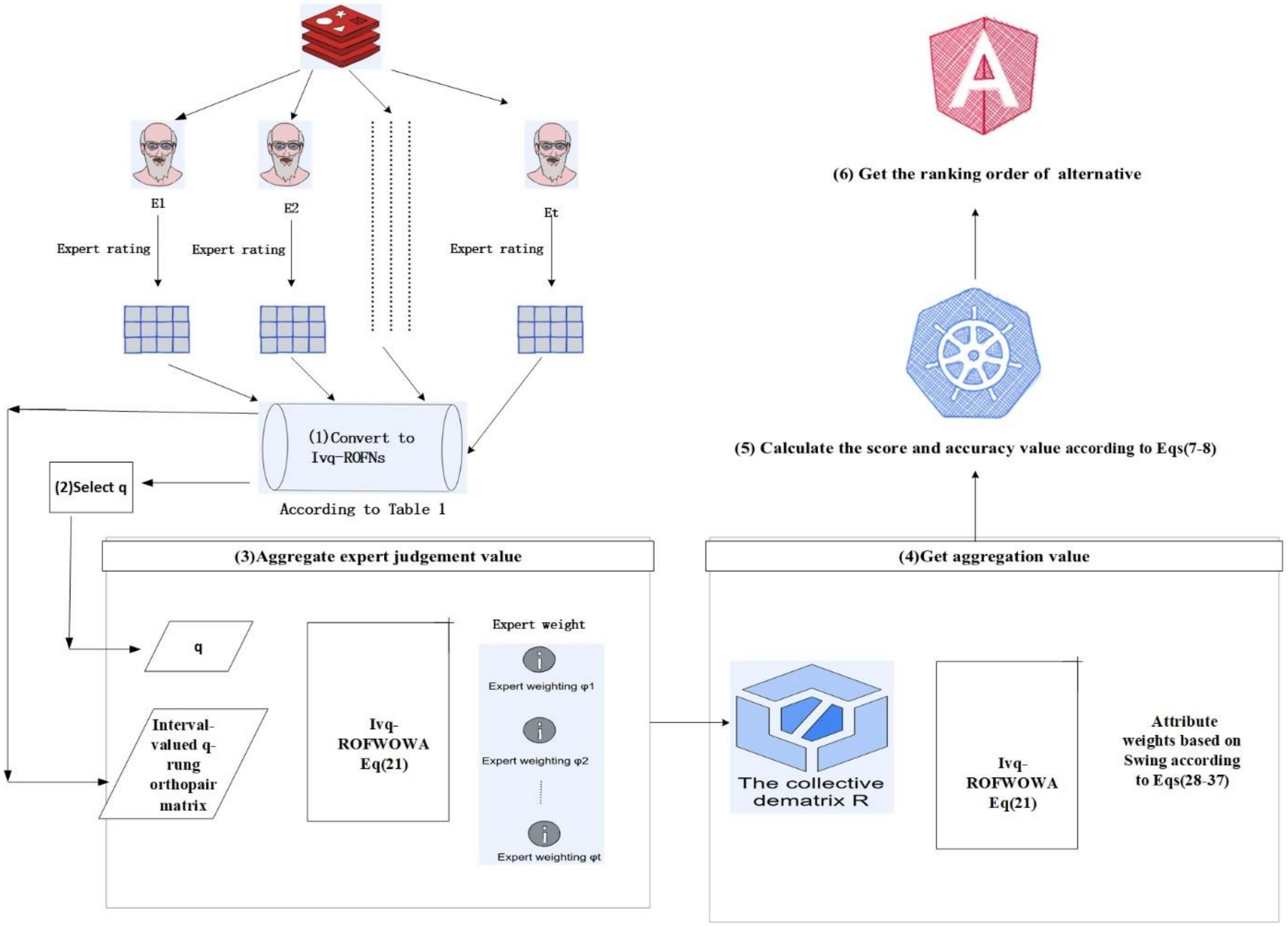

**Figure 3** MAGDM flow based on IVq-ROFWOWA and Swing method

# 4 Cases implementation

## 4.1 Learning effectiveness evaluation case

Learning effectiveness is essential for measuring how much learners have acquired knowledge, skills, and attitudes. Its evaluation can help educators better understand teaching effectiveness, guide learners in formulating learning strategies, and adjust teaching strategies, thus achieving the goals of education and teaching. In educational evaluation, learning effectiveness evaluation is a goal-oriented evaluation method that focuses on whether learners have achieved the expected learning objectives. During the evaluation process, many uncertain factors exist, such as individual differences among learners, the complexity of subject knowledge structures, etc. These factors may lead to uncertain fuzzy evaluation results. Our method can handle these uncertainties by establishing a fuzzy decision-making model to quantify the uncertainty and reduce the error of the evaluation results.

In the process of learning evaluation, five factors need to be evaluated: $c_1$ (learning interest), $c_2$ (learning style), $c_3$ (personality traits), $c_4$ (cognitive ability), and $c_5$ (external environment) with unknown weights but satisfying $\sum_{j=1}^{5} w_j = 1$, $w_j \geq 0$. Four experts were invited to conduct evaluations based on the learning evaluation model to evaluate the student's learning effectiveness. Their weight is vector $\varphi = (0.2445, 0.2494, 0.2488, 0.2573)$.

The evaluation results are classified into five levels: $x_1$ — Excellent, $x_2$ — s Good, $x_3$ — Fair, $x_4$ — Poor, and $x_5$ — Very Poor. **Tables 2-5** show evaluation matrix of their learning effectiveness level. Following steps are implemented.

**Table 2** Learning effectiveness level evaluation matrix given by Expert 1

| | $c_1$ | $c_2$ | $c_3$ | $c_4$ | $c_5$ |
|---|---|---|---|---|---|
| $x_1$ | AI | VHI | AAI | VHI | AI |
| $x_2$ | VHI | CHI | CHI | CHI | AAI |
| $x_3$ | VHI | CHI | AAI | HI | CHI |
| $x_4$ | CHI | AAI | AI | HI | AAI |
| $x_5$ | AI | AI | HI | AAI | HI |

**Table 3** Learning effectiveness level evaluation matrix given by Expert 2

| | $c_1$ | $c_2$ | $c_3$ | $c_4$ | $c_5$ |
|---|---|---|---|---|---|
| $x_1$ | CHI | VHI | HI | AAI | CHI |
| $x_2$ | VHI | HI | VHI | CHI | CHI |
| $x_3$ | VHI | VHI | HI | AAI | VHI |
| $x_4$ | AAI | BA | AAI | AAI | AAI |
| $x_5$ | AAI | HI | VHI | VHI | AI |

**Table 4** Learning effectiveness level evaluation matrix given by Expert 3

| | $c_1$ | $c_2$ | $c_3$ | $c_4$ | $c_5$ |
|---|---|---|---|---|---|
| $x_1$ | CHI | HI | AAI | AAI | AI |
| $x_2$ | CHI | VHI | CHI | AAI | HI |
| $x_3$ | VHI | HI | VHI | VHI | VHI |
| $x_4$ | AAI | BA | AAI | HI | AI |
| $x_5$ | AAI | AI | CHI | VHI | CHI |

**Table 5** Learning effectiveness level evaluation matrix given by Expert 4

| | $c_1$ | $c_2$ | $c_3$ | $c_4$ | $c_5$ |
|---|---|---|---|---|---|
| $x_1$ | CHI | HI | CHI | CHI | AAI |
| $x_2$ | HI | CHI | HI | VHI | CHI |
| $x_3$ | VHI | CHI | CHI | HI | AAI |
| $x_4$ | AI | CHI | HI | CHI | CHI |
| $x_5$ | AAI | AI | CHI | VHI | CHI |

(1) To enhance the interpretability of our fuzzy decision-making method and further improve the explanation and description of evaluation results, we convert the level matrix to a decision matrix as shown in **Tables 6-9**.

**Table 6** Decision matrix $A^{(1)}$ given by expert 1

| | $c_1$ | $c_2$ | $c_3$ | $c_4$ | $c_5$ |
|---|---|---|---|---|---|
| $x_1$ | ⟨[0.45,0.55], [0.45,0.55]⟩ | ⟨[0.80,0.90], [0.10,0.20]⟩ | ⟨[0.55,0.65], [0.35,0.45]⟩ | ⟨[0.80,0.90], [0.10,0.20]⟩ | ⟨[0.45,0.55], [0.45,0.55]⟩ |
| $x_2$ | ⟨[0.80,0.90], [0.10,0.20]⟩ | ⟨[0.90,0.95], [0.05,0.10]⟩ | ⟨[0.90,0.95], [0.05,0.10]⟩ | ⟨[0.90,0.95], [0.05,0.10]⟩ | ⟨[0.55,0.65], [0.35,0.45]⟩ |
| $x_3$ | ⟨[0.80,0.90], [0.10,0.20]⟩ | ⟨[0.90,0.95], [0.05,0.10]⟩ | ⟨[0.55,0.65], [0.35,0.45]⟩ | ⟨[0.65,0.80], [0.20,0.35]⟩ | ⟨[0.90,0.95], [0.05,0.10]⟩ |
| $x_4$ | ⟨[0.90,0.95], [0.05,0.10]⟩ | ⟨[0.55,0.65], [0.35,0.45]⟩ | ⟨[0.45,0.55], [0.45,0.55]⟩ | ⟨[0.65,0.80], [0.20,0.35]⟩ | ⟨[0.55,0.65], [0.35,0.45]⟩ |
| $x_5$ | ⟨[0.45,0.55], [0.45,0.55]⟩ | ⟨[0.45,0.55], [0.45,0.55]⟩ | ⟨[0.65,0.80], [0.20,0.35]⟩ | ⟨[0.55,0.65], [0.35,0.45]⟩ | ⟨[0.65,0.80], [0.20,0.35]⟩ |

**Table 7** Decision matrix $A^{(2)}$ given by expert 2

| | $c_1$ | $c_2$ | $c_3$ | $c_4$ | $c_5$ |
|---|---|---|---|---|---|
| $x_1$ | ⟨[0.90,0.95], [0.05,0.10]⟩ | ⟨[0.80,0.90], [0.10,0.20]⟩ | ⟨[0.65,0.80], [0.20,0.35]⟩ | ⟨[0.55,0.65], [0.35,0.45]⟩ | ⟨[0.90,0.95], [0.05,0.10]⟩ |
| $x_2$ | ⟨[0.80,0.90], [0.10,0.20]⟩ | ⟨[0.65,0.80], [0.20,0.35]⟩ | ⟨[0.80,0.90], [0.10,0.20]⟩ | ⟨[0.90,0.95], [0.05,0.10]⟩ | ⟨[0.90,0.95], [0.05,0.10]⟩ |
| $x_3$ | ⟨[0.80,0.90], [0.10,0.20]⟩ | ⟨[0.80,0.90], [0.10,0.20]⟩ | ⟨[0.65,0.80], [0.25,0.35]⟩ | ⟨[0.55,0.65], [0.35,0.45]⟩ | ⟨[0.80,0.90], [0.10,0.20]⟩ |
| $x_4$ | ⟨[0.55,0.65], [0.35,0.45]⟩ | ⟨[0.35,0.45], [0.55,0.65]⟩ | ⟨[0.55,0.65], [0.35,0.45]⟩ | ⟨[0.55,0.65], [0.35,0.45]⟩ | ⟨[0.55,0.65], [0.35,0.45]⟩ |
| $x_5$ | ⟨[0.55,0.65], [0.35,0.45]⟩ | ⟨[0.65,0.80], [0.20,0.35]⟩ | ⟨[0.80,0.90], [0.10,0.20]⟩ | ⟨[0.80,0.90], [0.10,0.20]⟩ | ⟨[0.45,0.55], [0.45,0.55]⟩ |

**Table 8** Decision matrix $A^{(3)}$ given by expert 3

| | $c_1$ | $c_2$ | $c_3$ | $c_4$ | $c_5$ |
|---|---|---|---|---|---|
| $x_1$ | ⟨[0.90,0.95], [0.05,0.10]⟩ | ⟨[0.65,0.80], [0.20,0.35]⟩ | ⟨[0.55,0.65], [0.35,0.45]⟩ | ⟨[0.55,0.65], [0.35,0.45]⟩ | ⟨[0.45,0.55], [0.45,0.55]⟩ |
| $x_2$ | ⟨[0.90,0.95], [0.05,0.10]⟩ | ⟨[0.80,0.90], [0.10,0.20]⟩ | ⟨[0.95,0.99], [0.01,0.05]⟩ | ⟨[0.55,0.65], [0.35,0.45]⟩ | ⟨[0.65,0.80], [0.20,0.35]⟩ |
| $x_3$ | ⟨[0.80,0.90], [0.10,0.20]⟩ | ⟨0.65,0.80], [0.20,0.35]⟩ | ⟨[0.80,0.90], [0.10,0.20]⟩ | ⟨[0.80,0.90], [0.10,0.20]⟩ | ⟨[0.80,0.90], [0.10,0.20]⟩ |
| $x_4$ | ⟨[0.55,0.65], [0.35,0.45]⟩ | ⟨[0.35,0.45], [0.55,0.65]⟩ | ⟨[0.55,0.65], [0.35,0.45]⟩ | ⟨[0.65,0.80], [0.20,0.35]⟩ | ⟨[0.45,0.55], [0.45,0.55]⟩ |
| $x_5$ | ⟨[0.55,0.65], [0.35,0.45]⟩ | ⟨[0.45,0.55], [0.45,0.55]⟩ | ⟨[0.90,0.95], [0.05,0.10]⟩ | ⟨[0.80,0.90], [0.10,0.20]⟩ | ⟨[0.90,0.95], [0.05,0.10]⟩ |

**Table 9** Decision matrix $A^{(4)}$ given by expert 4

| | $c_1$ | $c_2$ | $c_3$ | $c_4$ | $c_5$ |
|---|---|---|---|---|---|
| $x_1$ | ⟨[0.90,0.95], [0.05,0.10]⟩ | ⟨[0.65,0.80], [0.20,0.35]⟩ | ⟨[0.90,0.95], [0.05,0.10]⟩ | ⟨[0.90,0.95], [0.05,0.10]⟩ | ⟨[0.55,0.65, [0.35,0.45]⟩ |
| $x_2$ | ⟨[0.65,0.80], [0.20,0.35]⟩ | ⟨[0.90,0.95], [0.05,0.10]⟩ | ⟨[0.65,0.80], [0.20,0.35]⟩ | ⟨[0.80,0.90], [0.10,0.20]⟩ | ⟨[0.90,0.95], [0.05,0.10]⟩ |
| $x_3$ | ⟨[0.80,0.90], [0.10,0.20]⟩ | ⟨[0.90,0.95], [0.05,0.10]⟩ | ⟨[0.90,0.95], [0.05,0.10]⟩ | ⟨[0.65,0.80], [0.20,0.35]⟩ | ⟨[0.55,0.65], [0.35,0.45]⟩ |
| $x_4$ | ⟨[0.45,0.55], [0.45,0.55]⟩ | ⟨[0.90,0.95], [0.05,0.10]⟩ | ⟨[0.65,0.80], [0.20,0.35]⟩ | ⟨[0.90,0.95], [0.05,0.10]⟩ | ⟨[0.90,0.95], [0.05,0.10]⟩ |
| $x_5$ | ⟨[0.55,0.65, [0.35,0.45]⟩ | ⟨[0.45,0.55], [0.45,0.55]⟩ | ⟨[0.80,0.90], [0.10,0.20]⟩ | ⟨[0.80,0.90], [0.10,0.20]⟩ | ⟨[0.90,0.95], [0.05,0.10]⟩ |

(2) According to the observation, when $q = 2$, the elements in $A^{(k)}$ (k=1, 2, 3, 4) satisfy the condition defined by the IVq-ROFSs. The standard evaluation matrix $A^{(k)}$ is aggregated by using the IVq-ROFWOWA operator combined with the expert weight $\varphi = 0.2445, 0.2494, 0.2488, 0.2573)$. The aggregated matrix $R = (r_{ij})_{m\times n}$ can be obtained and shown in **Table 10**.

**Table 10** Aggregation matrix R

| | $c_1$ | $c_2$ | $c_3$ | $c_4$ | $c_5$ |
|---|---|---|---|---|---|
| $x_1$ | ⟨[0.78,0.84], [0.21,0.27]⟩ | ⟨[0.72,0.84], [0.15,0.28]⟩ | ⟨[0.66,0.75], [0.26,0.36]⟩ | ⟨[0.69,0.78], [0.24,0.32]⟩ | ⟨[0.58,0.67], [0.35,0.44]⟩ |
| $x_2$ | ⟨[0.78,0.88], [0.12,0.22]⟩ | ⟨[0.81,0.89], [0.11,0.21]⟩ | ⟨[0.81,0.89], [0.11,0.21]⟩ | ⟨[0.78,0.85], [0.18,0.24]⟩ | ⟨[0.74,0.83], [0.20,0.28]⟩ |
| $x_3$ | ⟨[0.80,0.90], [0.10,0.20]⟩ | ⟨[0.81,0.89], [0.11,0.21]⟩ | ⟨[0.72,0.82], [0.20,0.30]⟩ | ⟨[0.66,0.78], [0.22,0.34]⟩ | ⟨[0.76,0.84], [0.18,0.26]⟩ |
| $x_4$ | ⟨[0.61,0.69], [0.33,0.41]⟩ | ⟨[0.54,0.62], [0.41,0.49]⟩ | ⟨[0.55,0.66], [0.34,0.45]⟩ | ⟨[0.68,0.79], [0.22,0.33]⟩ | ⟨[0.61,0.69], [0.33,0.41]⟩ |
| $x_5$ | ⟨[0.52,0.62], [0.37,0.47]⟩ | ⟨[0.50,0.61], [0.39,0.50]⟩ | ⟨[0.78,0.88], [0.12,0.22]⟩ | ⟨[0.73,0.83], [0.19,0.27]⟩ | ⟨[0.72,0.80], [0.23,0.31]⟩ |

(3) Calculate the attribute weights according to Eqs. (28)-(37), we can obtain the attribute weight vector w as follows

$$\mathrm{w} = (0.1961,\ 0.1961,\ 0.1961,\ 0.1961,\ 0.2156).$$

(4) The aggregation matrix $R$ is aggregated using the IVq-ROFWOWA operator with the attribute weight vector $\mathrm{w} = (0.1961,\ 0.1961,\ 0.1961,\ 0.1961,\ 0.2156)$. The aggregating results of alternatives are

$$r_1 = ([0.6897, 0.7803], [0.2579, 0.3420]),$$
$$r_2 = ([0.7869, 0.8746], [0.1532, 0.2391]),$$
$$r_3 = ([0.7492, 0.8493], [0.1755, 0.2701]),$$
$$r_4 = ([0.6003, 0.6947], [0.3344, 0.4252]),$$
$$r_5 = ([0.6524, 0.7503], [0.2841, 0.3734]).$$

(5) The score function is used to calculate the scores of $r_i$ as $S(r_1) = 0.7252$, $S(r_2) = 0.8259$, $S(r_3) = 0.7947$, $S(r_4) = 0.6376$, $S(r_5) = 0.6921$. The ranking of the scores of each alternative is $S(r_3) > S(r_4) > S(r_2) > S(r_1) > S(r_5)$.

(6) According to the score ranking in step (5), the alternative ranking is

$$x_2 > x_3 > x_1 > x_5 > x_4,$$

and the optimal alternative is $x_2$. The student's level of learning achievement is $x_2$ and the student's learning performance is good, which is consistent with their actual performance in the classroom. The proposed method of evaluating learning achievement is thus feasible and effective.

## 4.2 Comparative analysis of operators

In this section, the Hamacher (**Darko and Liang, 2020**), Frank (Riaz 2022), OWA (Yager), and IVq-ROFWOWA operators are used for comparative analysis. $\alpha = 2$ is set for the Frank operator, $\varphi = 2$ is set for the Hamacher operator, and $\lambda = 2$ is set for the IVq-ROFWOWA operator. By taking the values of $q$ from the interval [2,9], the scores and ranking of each learning effectiveness rating can be obtained, as shown in **Table 11 and Figure 4**.

**Table 11** Learning effectiveness level score and ranking of each operator

| | **IVQ-FWOWA** | **Hamacher** | **Frank** | **OWA** | Sort | |
|---|---|---|---|---|---|---|
| q=2 | $sf_1 = 0.7252$ | $S_1 = 0.7909$ | $C_1 = 0.7953$ | $U_1 = 0.8002$ | Hamacher | $x_2 > x_3 > x_1 > x_5 > x_4$ |
| | $sf_2 = 0.8259$ | $S_2 = 0.8606$ | $C_2 = 0.8623$ | $U_2 = 0.8643$ | Frank | $x_2 > x_3 > x_1 > x_5 > x_4$ |
| | $sf_3 = 0.7947$ | $S_3 = 0.8259$ | $C_3 = 0.8276$ | $U_3 = 0.8297$ | | |
| | $sf_4 = 0.6376$ | $S_4 = 0.7027$ | $C_4 = 0.7088$ | $U_4 = 0.7156$ | OWA | $x_2 > x_3 > x_1 > x_5 > x_4$ |
| | $sf_5 = 0.6921$ | $S_5 = 0.7513$ | $C_5 = 0.7559$ | $U_5 = 0.7610$ | FWOWA | $\boldsymbol{x_2 > x_3 > x_1 > x_5 > x_4}$ |
| q=3 | $sf_1 = 0.6907$ | $S_1 = 0.7511$ | $C_1 = 0.7575$ | $U_1 = 0.7634$ | Hamacher | $x_2 > x_3 > x_1 > x_4 > x_5$ |
| | $sf_2 = 0.7871$ | $S_2 = 0.8212$ | $C_2 = 0.8239$ | $U_2 = 0.8267$ | Frank | $x_2 > x_3 > x_1 > x_5 > x_4$ |
| | $sf_3 = 0.7540$ | $S_3 = 0.7838$ | $C_3 = 0.7866$ | $U_3 = 0.7893$ | | |
| | $sf_4 = 0.6126$ | $S_4 = 0.6714$ | $C_4 = 0.6790$ | $U_4 = 0.6862$ | OWA | $x_2 > x_3 > x_1 > x_5 > x_4$ |
| | $sf_5 = 0.6601$ | $S_5 = 0.7137$ | $C_5 = 0.7199$ | $U_5 = 0.7255$ | FWOWA | $\boldsymbol{x_2 > x_3 > x_1 > x_5 > x_4}$ |
| q=4 | $sf_1 = 0.6612$ | $S_1 = 0.7128$ | $C_1 = 0.7202$ | $U_1 = 0.7265$ | Hamacher | $x_2 > x_3 > x_1 > x_5 > x_4$ |
| | $sf_2 = 0.7506$ | $S_2 = 0.7828$ | $C_2 = 0.7864$ | $U_2 = 0.7897$ | Frank | $x_2 > x_3 > x_1 > x_5 > x_4$ |
| | $sf_3 = 0.7157$ | $S_3 = 0.7425$ | $C_3 = 0.7460$ | $U_3 = 0.7492$ | | |
| | $sf_4 = 0.5920$ | $S_4 = 0.6398$ | $C_4 = 0.6481$ | $U_4 = 0.6551$ | OWA | $x_2 > x_3 > x_1 > x_5 > x_4$ |
| | $sf_5 = 0.6329$ | $S_5 = 0.6767$ | $C_5 = 0.6836$ | $U_5 = 0.6892$ | FWOWA | $\boldsymbol{x_2 > x_3 > x_1 > x_5 > x_4}$ |
| q=5 | $sf_1 = 0.6381$ | $S_1 = 0.6819$ | $C_1 = 0.6897$ | $U_1 = 0.6958$ | Hamacher | $x_2 > x_3 > x_1 > x_5 > x_4$ |
| | $sf_2 = 0.7198$ | $S_2 = 0.7499$ | $C_2 = 0.7541$ | $U_2 = 0.7576$ | Frank | $x_2 > x_3 > x_1 > x_5 > x_4$ |
| | $sf_3 = 0.6840$ | $S_3 = 0.7080$ | $C_3 = 0.7119$ | $U_3 = 0.7151$ | | |
| | $sf_4 = 0.5763$ | $S_4 = 0.6149$ | $C_4 = 0.6232$ | $U_4 = 0.6298$ | OWA | $x_2 > x_3 > x_1 > x_5 > x_4$ |
| | $sf_5 = 0.6117$ | $y_5 = 0.6472$ | $C_5 = 0.6540$ | $U_5 = 0.6593$ | FWOWA | $\boldsymbol{x_2 > x_3 > x_1 > x_5 > x_4}$ |
| q=6 | $sf_1 = 0.6201$ | $S_1 = 0.6576$ | $C_1 = 0.6653$ | $U_1 = 0.6710$ | Hamacher | $x_2 > x_3 > x_1 > x_5 > x_4$ |
| | $sf_2 = 0.6944$ | $S_2 = 0.7222$ | $C_2 = 0.7267$ | $U_2 = 0.7301$ | Frank | $x_2 > x_3 > x_1 > x_5 > x_4$ |
| | $sf_3 = 0.6583$ | $S_3 = 0.6797$ | $C_3 = 0.6837$ | $U_3 = 0.6868$ | | |
| | $sf_4 = 0.5647$ | $S_4 = 0.5963$ | $C_4 = 0.6042$ | $U_4 = 0.6101$ | OWA | $x_2 > x_3 > x_1 > x_5 > x_4$ |
| | $sf_5 = 0.5954$ | $S_5 = 0.6243$ | $C_5 = 0.6308$ | $U_5 = 0.6356$ | FWOWA | $\boldsymbol{x_2 > x_3 > x_1 > x_5 > x_4}$ |
| q=7 | $sf_1 = 0.6060$ | $S_1 = 0.6382$ | $C_1 = 0.6455$ | $U_1 = 0.6508$ | Hamacher | $x_2 > x_3 > x_1 > x_5 > x_4$ |
| | $sf_2 = 0.6733$ | $S_2 = 0.6988$ | $C_2 = 0.7032$ | $U_2 = 0.7066$ | Frank | $x_2 > x_3 > x_1 > x_5 > x_4$ |
| | $sf_3 = 0.6374$ | $S_3 = 0.6565$ | $C_3 = 0.6604$ | $U_3 = 0.6634$ | | |
| | $sf_4 = 0.5560$ | $S_4 = 0.5823$ | $C_4 = 0.5896$ | $U_4 = 0.5949$ | OWA | $x_2 > x_3 > x_1 > x_5 > x_4$ |
| | $sf_5 = 0.5826$ | $S_5 = 0.6065$ | $C_5 = 0.6125$ | $U_5 = 0.6167$ | FWOWA | $\boldsymbol{x_2 > x_3 > x_1 > x_5 > x_4}$ |
| q=8 | $sf_1 = 0.5946$ | $S_1 = 0.6225$ | $C_1 = 0.6293$ | $U_1 = 0.6340$ | Hamacher | $x_2 > x_3 > x_1 > x_5 > x_4$ |
| | $sf_2 = 0.6556$ | $S_2 = 0.6787$ | $C_2 = 0.6831$ | $U_2 = 0.6863$ | Frank | $x_2 > x_3 > x_1 > x_5 > x_4$ |
| | $sf_3 = 0.6203$ | $S_3 = 0.6372$ | $C_3 = 0.6410$ | $U_3 = 0.6437$ | | |
| | $sf_4 = 0.5494$ | $S_4 = 0.5717$ | $C_4 = 0.5782$ | $U_4 = 0.5829$ | OWA | $x_2 > x_3 > x_1 > x_5 > x_4$ |
| | $sf_5 = 0.5724$ | $S_5 = 0.5924$ | $C_5 = 0.5978$ | $U_5 = 0.6015$ | FWOWA | $\boldsymbol{x_2 > x_3 > x_1 > x_5 > x_4}$ |
| q=9 | $sf_1 = 0.5853$ | $S_1 = 0.6095$ | $C_1 = 0.6157$ | $U_1 = 0.6199$ | Hamacher | $x_2 > x_3 > x_1 > x_5 > x_4$ |
| | $sf_2 = 0.6405$ | $S_2 = 0.6613$ | $C_2 = 0.6656$ | $U_2 = 0.6686$ | Frank | $x_2 > x_3 > x_1 > x_5 > x_4$ |
| | $sf_3 = 0.6061$ | $S_3 = 0.6210$ | $C_3 = 0.6246$ | $U_3 = 0.6271$ | OWA | $x_2 > x_3 > x_1 > x_5 > x_4$ |
| | $sf_4 = 0.5442$ | $S_4 = 0.5633$ | $C_4 = 0.5692$ | $U_4 = 0.5732$ | **FWOWA** | $\boldsymbol{x_2 > x_3 > x_1 > x_5 > x_4}$ |
| | $sf_5 = 0.5640$ | $S_5 = 0.5810$ | $C_5 = 0.5858$ | $U_5 = 0.5890$ | | |

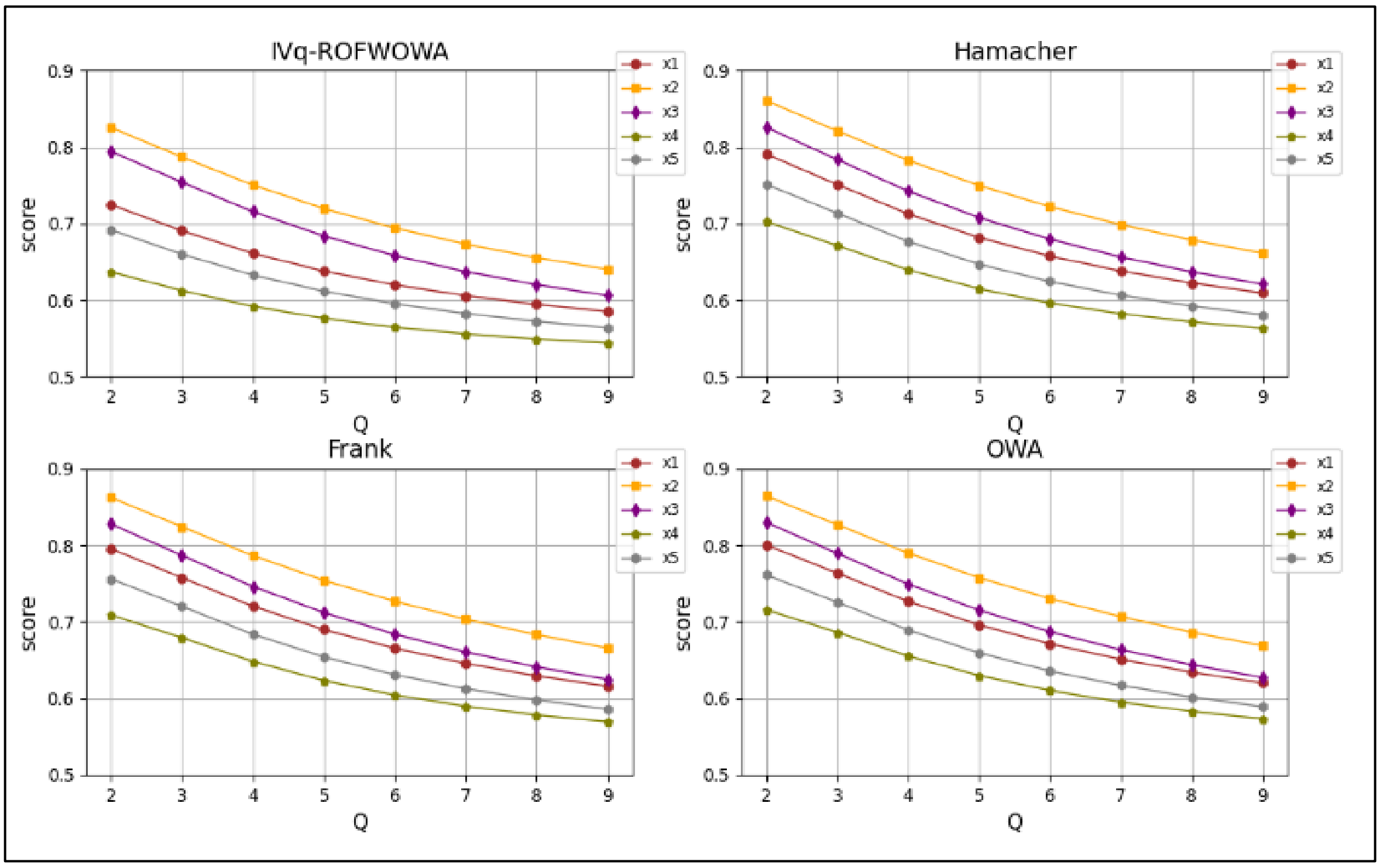

**Figure 4** Comparative analysis of operators

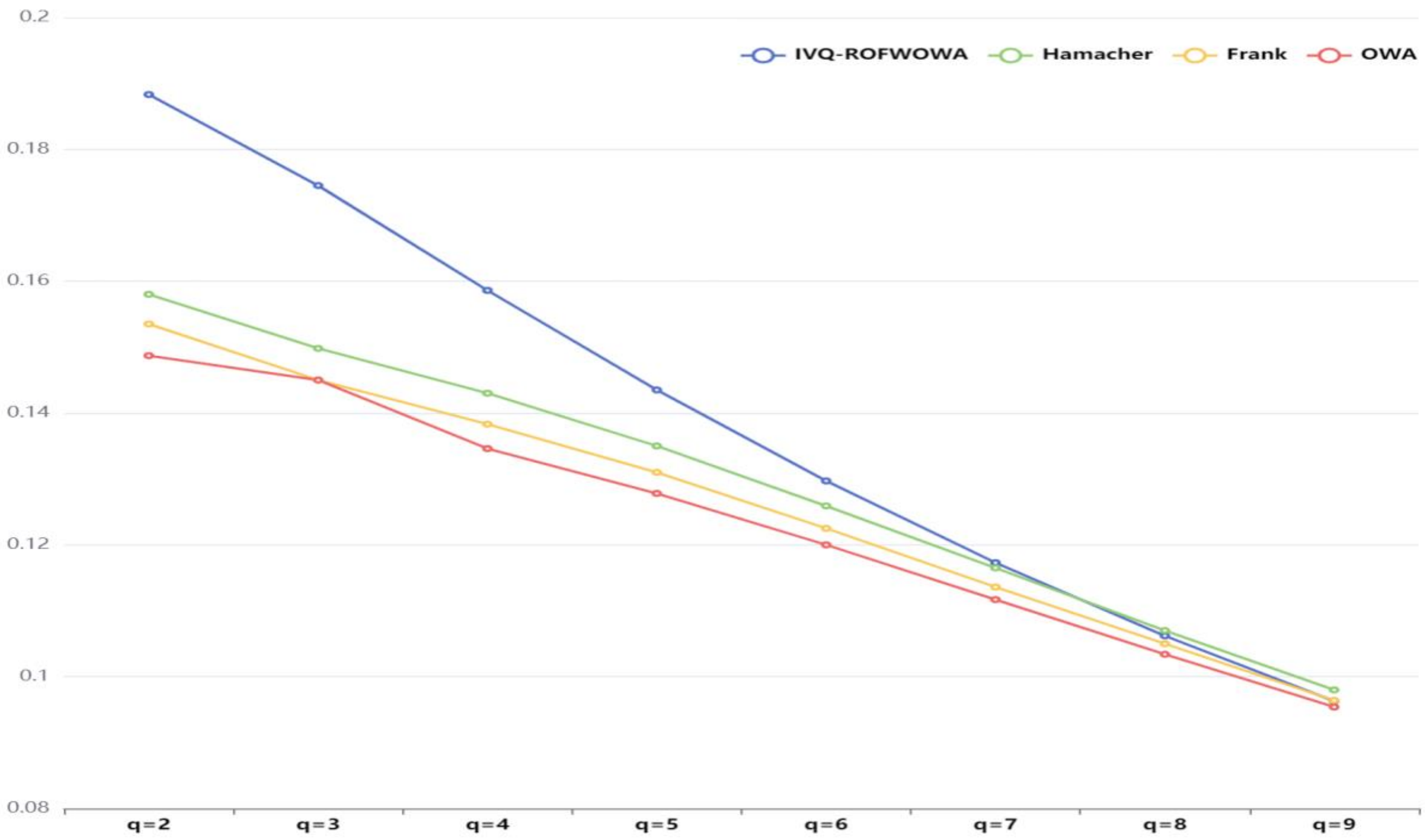

**Figure 5** Comparative analysis of operators

It can be seen from **Table 11 and Figure 4** that the IVq-ROFWOWA operator proposed has the same ranking as the Hamacher, Frank, and OWA operators under IVq-ROFSs. The ranking results of each alternative remain unchanged as $q$ changes. Moreover, we will arrange the scores of the schemes in descending order, then subtract the differences successively from front to back, calculate the cumulative sum of the differences, and finally take the average. **Figure 5** shows that the IVq-ROFWOWA operator has a more obvious difference in scores compared with Hamacher, Frank, and OWA operators. Thus, the operator we proposed is advantageous in the case.

## 4.3 Comparative analysis of attribute weights

In order to verify the Swing, the Project (Zhang et al., 2019) and MABAC (Verma, 2021) methods are used to compare with the Swing method in MAGDM as determining attribute weights. We will outline the main implementation steps.

The Projection method is a decision analysis technique used to determine weights based on the contribution of each criterion in a comprehensive evaluation (Zhang et al., 2019). The method involves projecting the evaluation criteria onto a weight vector to determine the weight of each criterion as shown in Eq. (40).

$$\mathrm{Proj}_{Y^*}(Y_k) = \frac{\sum_{i=1}^{m}\sum_{j=1}^{n} y_{ij}^{(k)} y_{ij}^*}{\sqrt{\sum_{i=1}^{m}\sum_{j=1}^{n} {y_{ij}^*}^2}}. \tag{40}$$

In Eq. (40), $\mathrm{Proj}_{Y^*}(Y_k)$ presents projection value. The larger the projection value of $Y_\mathrm{k}$, the $Y_\mathrm{k}$ is closer to the $Y^*$, and the k-th attribute weight is larger.

In order to get the weights vector $w_{Projection} = (w_1, w_2, \ldots, w_n)$ by these projections, we need to make the following transformation

$$w_i = \frac{Proj_{Y^*}(Y_k)}{\sum_{\mathrm{i}=1}^{\mathrm{n}} Proj_{Y^*}(Y_k)}. \tag{41}$$

The MABAC method establishes a comparison benchmark firstly by determining the ideal solutions, which utilizes the concept of boundary approximation to assess the degree of proximity. The MABAC method is then used to solve the attribute weights in MAGDM (Verma, 2021), which are derived by the following steps.

**Step 1.** Eqs. (42) and (43) are used to normalize the positive and negative attributes of the decision matrix, respectively.

$$r_{ij}^{*}=\frac{r_{ij}-r_{i}^{-}}{r_{i}^{+}-r_{i}^{-}};\mathrm{i}=1,\ldots,\mathrm{m},\ \mathrm{j}=1,\ldots,\mathrm{n}. \tag{42}$$

$$r_{ij}^{*}=\frac{r_{ij}-r_{i}^{+}}{r_{i}^{-}-r_{i}^{+}};i=1,\ldots,m,\ j=1,\ldots,n, \tag{43}$$

where $r_{ij}^{*}$ indicates the normalized value of the decision matrix of i-th alternative in j-th attribute.

Additionally, we have $r_i^{+}=\max(\mathrm{r}_1,\mathrm{r}_2,\ldots,\mathrm{r}_\mathrm{m})$ and $r_i^{-}=min(r_1,r_2,\ldots,r_m)$.

(1) Given the normalized values of the decision matrix, the weighted normalized values of each attribute are obtained from Eq. (44):

$$\widehat{r_{ij}}=r_{ij}^{*}w_j^{*};i=1,\ldots.m,\ \ j=1,\ldots,n. \tag{44}$$

(2) The values of the border approximation area matrix are obtained in Eq. (45).

$$g_j=\left(\prod_{i=1}^{m}\widehat{r_{ij}}\right)^{\frac{1}{m}},\ \ j=1,\ldots.,n. \tag{45}$$

By determining the importance of the border approximation area matrix, a $nx1$ matrix can be obtained.

(3) With respect to the amounts of the border approximation area matrix and the weighted normalized values of each attribute, the distance of the alternatives from the border approximation area is determined as

$$q_{ij}=\widehat{r_{ij}}-g_j,\ \ i=1,\ldots.m,\ \ j=1,\ldots.,n. \tag{46}$$

(4) The weight vector $w_{MABAC}=(w_1,w_2,\ldots,w_n)$ can be solved by

$$w_i=\frac{\sum_{i=1}^{n}\mathrm{q_{ij}}}{\sum_{i=1}^{i=n}\sum_{i=1}^{i=n}\mathrm{q_{ij}}}. \tag{47}$$

According to Eqs. (37)，(41) and (47), we can obtain their attribute weights as

$$w_{MABAC}=(0.2001,\ 0.2017,\ 0.2014,\ 0.0.1979,\ 0.1989),$$

$$w_{Projection}=(0.1974,\ 0.2239,\ 0.2159,.0.1732,\ 0.1896),$$

$$w_{Swing}=(0.1961,\ 0.1961,\ 0.1961,\ 0.1961,\ 0.2156).$$

We replace attribute weights vector $w_{Swing}$ with $w_{MABAC}$、$w_{Projection}$ in the case of section 4.1, respectively. The alternatives ranking results are the same. And we can’t select the best attribute weight methods from Swing, MABAC, and Projection. In order to compare Swing, Projection, and MABAC methods, we adopt a competitiveness theory to assess the strength of competition among different strategies. As seen in Eq. (48) (Kvålseth 2022), a larger value of S indicates the formation of a monopoly by the solution

$$\mathrm{S} = \sum_{j=1}^{j=n}\left(\frac{S(r_j)}{\sum_{i=1}^{i=n} S(r_j)}\right)^2, \tag{48}$$

where $S(r_j)$ is the score function.

By Eq. (48), the competition values $\mathrm{S}_{\mathrm{Swing}}$, $\mathrm{S}_{\mathrm{MABAC}}$ and $\mathrm{S}_{\mathrm{Projection}}$ of Swing， MABAC and Projection are obtained, respectively. And we can calculate $\mathrm{S}_{\mathrm{Swing}}$, $\mathrm{S}_{\mathrm{MABAC}}$ and $\mathrm{S}_{\mathrm{Projection}}$ when $q$ changes from 2 to 9. The value $(\mathrm{S}_{\mathrm{Swing}} - \mathrm{S}_{\mathrm{MABAC}}) \times 1000$ and $(\mathrm{S}_{\mathrm{Projection}} - \mathrm{S}_{\mathrm{MABAC}}) \times 1000$ also are calculated when $q$ changes from 2 to 9. As shown in **Figure 6**, the bars indicate the competitive values of the corresponding methods, and the line represents the comparative relationship between competitive values of $\mathrm{S}_{\mathrm{Swing}} - \mathrm{S}_{\mathrm{MABAC}}$ and $\mathrm{S}_{\mathrm{Projection}} - \mathrm{S}_{\mathrm{MABAC}}$.

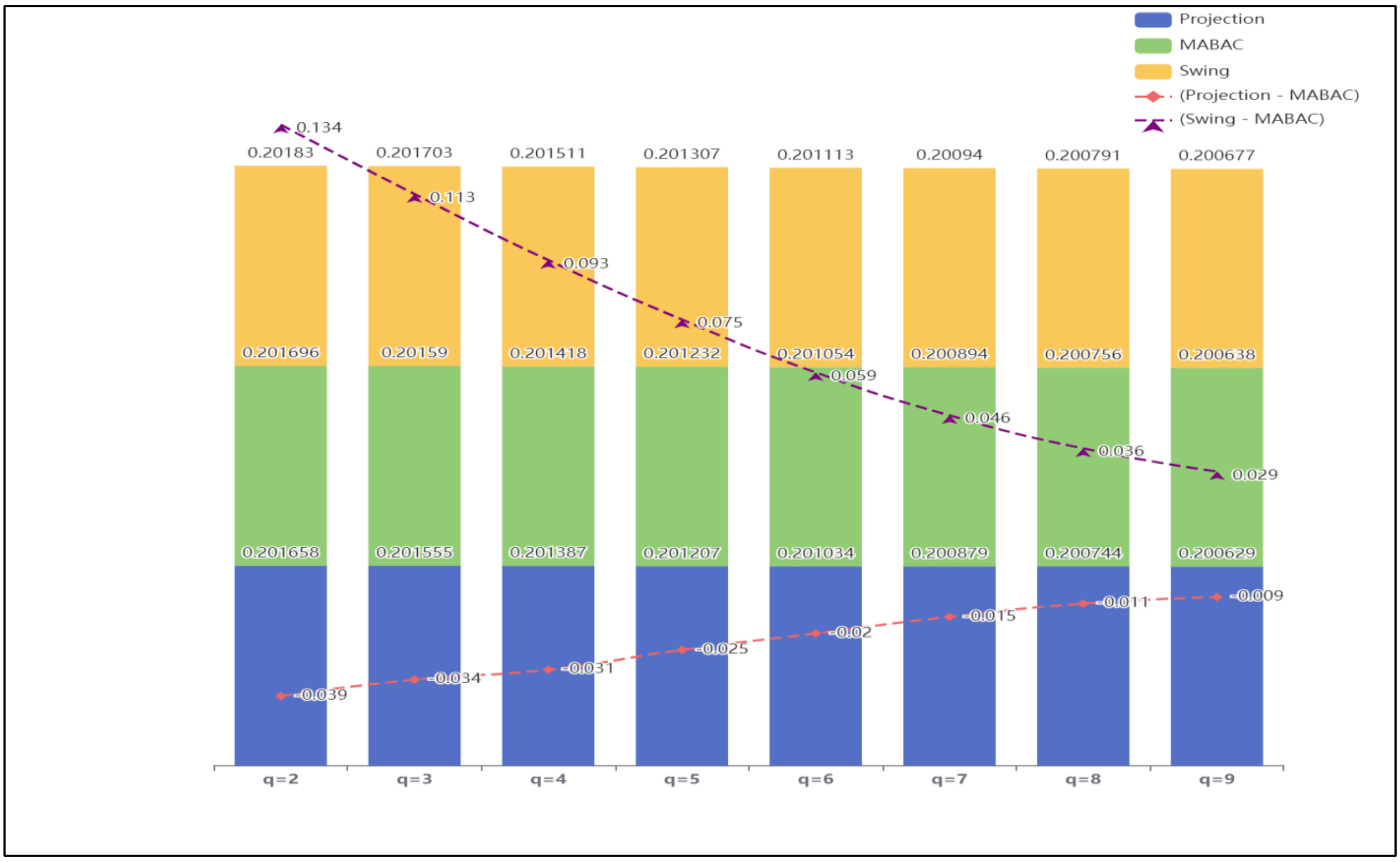

**Figure 6** Sensitivity analysis of competitiveness

From **Figure 6**, it can be observed that $q$ increases from 2 to 9, the competition values of attribute weights decrease. However, the line of $S_{Swing} - S_{MABAC}$ consistently has values greater than 0, which indicates that Swing is more effective than MABAC. While the line of $S_{Projection} - S_{MABAC}$ has values consistently smaller than 0, which indicates that the Projection exhibits lower performance compared to MABAC. Therefore, we can conclude that the Swing method is important to more effective than MABAC and Projection methods in deriving attribute weights in the case in section 4.1.

## 5 Conclusion

In this paper, A Swing-based MAGDM method is introduced, which use weber operator under IVq-ROFSs, and which expands the horizons of learning effectiveness evaluation. The extended IVq-ROFWOWA operator not only has excellent scaling ability but also enables experts to express judgment values more freely. Besides, the Swing algorithm enhances better differentiation in ranking alternatives compared to other methods. The implementation results of the learning effectiveness evaluation case demonstrate that the proposed Swing-based MAGDM method is consistent with the actual situation and produces results consistent with those obtained by existing expert, confirming the feasibility and effectiveness of the proposed evaluation method.

However, the learning effectiveness evaluation model developed in this paper involves several experts evaluating a few students. Adaptive and automated learning effectiveness evaluation is still an area for further research. Regarding future developments, we plan to focus on the learning effectiveness evaluation model by combining artificial intelligence and extensive data methods.

**Acknowledgment:** This paper has received support from the Ministry of Education Humanities and Social Sciences Planning Fund Project under Grant 22YJA880051, as well as partial funding from the Department of Education of Jiangxi Province, China, under Grant GJJ2200535and 22YB052.

**Data Availability Statement:** This paper contains the data was used to support and validate the results of this research.

**Conflicts of Interest:** These authors declare that there are no competing interests among them.

## References

1. Alamri, H. A., Watson, S., Watson, W. (2021). Learning technology models that support personalization within blended learning environments in higher education[J]. TechTrends, 65, 62-78. https://doi.org/ 10.1007/s11528-020-00530-3.
2. Araghi, T., Busch, C. A., Cooper, K. M. (2023). The aspects of active-learning science courses that exacerbate and alleviate depression in undergraduates[J]. CBE—Life Sciences Education, 22(2), ar26. https://doi.org/10.1187/cbe.22-10-0199.
3. Asif, K., Jamil, M. K., Karamti, H., Azeem, M., Ullah, K. (2023). Randić energies for T-spherical fuzzy Hamacher graphs and their applications in decision making for business plans[J]. Computational and Applied Mathematics, 42(3), 106.
4. Atanassov, K. T., Atanassov, K. T. (1999). Interval valued intuitionistic fuzzy sets[J]. Intuitionistic Fuzzy Sets: Theory and Applications, 139-177.
5. Csapó, B., Molnár, G. (2019). Online diagnostic assessment in support of personalized teaching and learning: The eDia system[J]. Frontiers in psychology, 10, 1522. https://doi.org/10.3389/fpsyg.2019.01522.
6. Culver, C. (2022). Learning as a peer assessor: evaluating peer-assessment strategies. Assessment & Evaluation in Higher Education, 1-17. https://doi.org/10.1080/02602938.2022.2107167.
7. Frank M.J. (1979). On the simultaneous associativity of F (x, y) and x+y−F(x, y) [J]. Aequat. math, 18,266-267. https://doi.org/10.1007/BF01844082.
8. Hamacher, H. (1975). Über logische Verknüpfungen unscharfer Aussagen und deren zugehörige Bewertungsfunktionen.
9. Jan, A. U., Barukab, O., Khan, A., Jun, Y. B., Khan, S. A. (2023). Cubical fuzzy Hamacher aggregation operators in multi-attribute decision-making problems[J]. Computational and Applied Mathematics, 42(3), 1-25.
10. Javed, M., Javeed, S., Ahmad, J., Ullah, K., Zedam, L. (2022). Approach to multiattribute decision-making problems based on neutrality aggregation operators of picture fuzzy information[J]. Journal of Function Spaces, 2022. https://doi.org/10.1155/2022/2762067.
11. Joshi, B. P., Singh, A., Bhatt, P. K., Vaisla, K. S. (2018). Interval valued q-rung orthopair fuzzy sets and their properties[J]. Journal of Intelligent & Fuzzy Systems, 35(5), 5225-5230. https://doi.org/10.3233/JIFS-169806.
12. Kvålseth, T. O. (2022). Cautionary Note About the Herfindahl-Hirschman Index of Market (Industry) Concentration[J]. Contemporary Economics, 16(1), 51-60.
13. Laktionov, I., Vovna, O., Kabanets, M. (2023). Computer-Oriented Method of Adaptive Monitoring and Control of Temperature and Humidity Mode of Greenhouse Production. Baltic Journal of Modern Computing, 11(1). https://doi.org/10.22364/bjmc.2023.11.1.12
14. Luqman, A., Shahzadi, G. (2023). Multi-criteria group decision-making based on the interval-valued q-rung orthopair fuzzy SIR approach for green supply chain evaluation and selection. Granular Computing, 1-18.

15. Paśko, Ł., Mądziel, M., Stadnicka, D., Dec, G., Carreras-Coch, A., Solé-Beteta, X., ... Atzeni, D. (2022). Plan and Develop Advanced Knowledge and Skills for Future Industrial Employees in the Field of Artificial Intelligence, Internet of Things and Edge Computing[J]. Sustainability, 14(6), 3312. https://doi.org/10.3390/su14063312.

16. Riaz, M., Farid, H. M. A., Wang, W., Pamucar, D. (2022). Interval-valued linear Diophantine fuzzy Frank aggregation operators with multi-criteria decision-making[J]. Mathematics, 10(11), 1811. https://doi.org/10.3390/math10111811.

17. Seker, S., Bağlan, F. B., Aydin, N., Deveci, M., Ding, W. (2023). Risk assessment approach for analyzing risk factors to overcome pandemic using interval-valued q-rung orthopair fuzzy decision making method[J]. Applied Soft Computing, 132, 109891. https://doi.org/10.1016/j.asoc.2022.109891.

18. Verma, R. (2021). Fuzzy MABAC method based on new exponential fuzzy information measures. Soft Computing, 25(14), 9575-9589.

19. Wan, B., Hu, Z., Garg, H., Cheng, Y., Han, M. (2023). An integrated group decision-making method for the evaluation of hypertension follow-up systems using interval-valued q-rung orthopair fuzzy sets. Complex & Intelligent Systems, 1-34. https://doi.org/10.1007/s40747-022-00953-w.

20. Wang, J., Gao, H., Wei, G., Wei, Y. (2019). Methods for multiple-attribute group decision making with q-rung interval-valued orthopair fuzzy information and their applications to the selection of green suppliers[J]. Symmetry, 11(1), 56. https://doi.org/10.3390/sym11010056.

21. Wang, M., Lv, Z. (2022). Construction of personalized learning and knowledge system of chemistry specialty via the internet of things and clustering algorithm[J]. The Journal of Supercomputing, 78(8), 10997-11014. https://doi.org/10.1007/s11227-022-04315-8.

22. Weber, S. (1983). A general concept of fuzzy connectives, negations and implications based on t-norms and t-conorms[J]. Fuzzy sets and systems, 11(1-3), 115-134. https://doi.org/10.1016/S0165-0114(83)80073-6.

23. Wu, T., Wang, Y., Zhang, J. (2023). Research on multilayer fast equalization strategy of Li-ion battery based on adaptive neural fuzzy inference system. Journal of Energy Storage, 67, 107574. https://doi.org/10.1016/j.est.2023.107574

24. Yager, R. R. (1988). On ordered weighted averaging aggregation operators in multicriteria decisionmaking[J]. IEEE Transactions on systems, Man, and Cybernetics, 18(1), 183-190. https://doi.org/10.1109/21.87068.

25. Yager, R. R. (2016). Generalized orthopair fuzzy sets[J]. IEEE Transactions on Fuzzy Systems, 25(5), 1222-1230. https://doi.org/10.1109/TFUZZ.2016.2604005.

26. Yang, X., Zhu, Y., Zhang, Y., Wang, X., Yuan, Q. (2020). Large scale product graph construction for recommendation in e-commerce. arXiv preprint arXiv:2010.05525. https://doi.org/10.48550/arXiv.2010.05525.

27. Ye, J., Du, S., Yong, R. (2023). Hybrid weighted arithmetic and geometric averaging operator of cubic Z-numbers and its decision-making method. Journal of Control and Decision, 1-12.

28. Zhang, X., Gou, X., Xu, Z., & Liao, H. (2019). A projection method for multiple attribute group

decision making with probabilistic linguistic term sets. International Journal of Machine Learning and Cybernetics, 10, 2515-2528. https://doi.org/10.1007/s13042-018-0886-6

29. Zhong, Y., Zhang, H., Cao, L., Li, Y., Qin, Y., Luo, X. (2023). Power Muirhead mean operators of interval-valued intuitionistic fuzzy values in the framework of Dempster–Shafer theory for multiple criteria decision-making[J]. Soft Computing, 27(2), 763-782.

30. Zhou, J., Ran, F., Li, G., Peng, J., Li, K., Wang, Z. (2022). Classroom Learning Status Assessment Based on Deep Learning. Mathematical Problems in Engineering, 2022, 1-9. https://doi.org/10.1155/2022/7049458.